\documentclass[12pt,letterpaper]{iopart}
\usepackage{epsfig}
\usepackage{cite,psfig}
\usepackage{psfrag}
\usepackage{latexsym}
\usepackage{axodraw}
\usepackage{xspace}
\usepackage{amssymb}

\begin{document}

\def\be{\begin{equation}}
\def\ee{\end{equation}}
\def\bea{\begin{eqnarray}}
\def\eea{\end{eqnarray}}
\def\nub{\bar{\nu}}
\def\A{{\cal{A}}}
\def\N{{\cal{N}}}
\def\mnu{m_{\nu}}
\def\mnui{m_{\nu_i}}
\def\m0{m_{\nu_{0,i}}}
\def\Tnu{T_{\nu}}
\def\T0{T_{\nu_0}}
\def\mrel{m_{\nu_{\rm Rel}}(z)}
\def\mnr{m_{\nu_{\rm NR}}(z)}
\def\Veff{V_{\rm eff}}
\def\rhoi{\rho_{\nu_i}}
\def\eV{{\rm eV}}
\def\res{{\rm{res}}}
\def\max{{\rm{max}}}

\newcommand{\nubar}
{{\bar{\nu}}}
\newcommand{\CnuB}{C\(\nu\)B\xspace}
\newcommand{\CMB}{CMB\xspace}
\newcommand{\UHEnu}{EHEC\(\nu\)\xspace}
\newcommand{\lwig}{\mbox{\;\raisebox{.3ex}
    {$<$}$\!\!\!\!\!$\raisebox{-.9ex}{$\sim$}\;}}
\newcommand{\gwig}{\mbox{\;\raisebox{.3ex}
    {$>$}$\!\!\!\!\!$\raisebox{-.9ex}{$\sim$}}\;}

\begin{flushright}
{\large \tt DESY-06-088}
\end{flushright}

\title{Probing Neutrino Dark Energy with Extremely High-Energy Cosmic Neutrinos}

\author{Andreas~Ringwald and Lily~Schrempp}
\address{Deutsches Elektronen-Synchrotron DESY, Notkestra\ss e  85, 22607 Hamburg, Germany}

\ead{\mailto{andreas.ringwald@desy.de},
\mailto{lily.schrempp@desy.de}}

\begin{abstract}

Recently, a new non-Standard Model neutrino interaction mediated by a light scalar field was proposed, 
which renders the big-bang relic neutrinos of the cosmic neutrino background a natural dark energy candidate, the so-called Neutrino Dark Energy. As a further consequence of this 
interaction, the neutrino masses become functions of the neutrino energy densities and are thus promoted to dynamical, time/redshift dependent quantities. Such a possible neutrino mass 
variation introduces a redshift dependence into the resonance energies associated with the annihilation 
of extremely high-energy cosmic neutrinos on relic anti-neutrinos and vice versa into Z-bosons. In general, this annihilation process is expected to lead to 
sizeable absorption dips in the spectra to be observed on earth by neutrino observatories operating in the relevant energy region above 
$10^{13}$~GeV. In our analysis, we contrast the characteristic absorption features produced by constant and 
varying neutrino masses, including all thermal background effects caused by the relic neutrino motion. We firstly consider neutrinos from astrophysical sources and secondly neutrinos originating from the decomposition of topological defects using the appropriate fragmentation functions. 
On the one hand, independent of the nature of neutrino masses, our results illustrate the discovery potential 
for the cosmic neutrino background by means of relic neutrino absorption spectroscopy. On the other hand, 
they allow to estimate the prospects for testing its possible interpretation as source of 
Neutrino Dark Energy within the next decade by the neutrino observatories ANITA and LOFAR. 
\end{abstract}
\maketitle

\section{Introduction}

According to Big Bang Cosmology, in an expanding universe the freeze-out of a particle species occurs, when its interaction strength is too small to keep it in thermal equilibrium. Neutrinos, being the particles with the weakest known interactions, therefore are assumed to have already decoupled when the universe was just $\approx 1$~ s old, thereby guaranteeing a substantial relic neutrino abundance today with average number density $n_{\nu_{0,i}}=n_{\nubar_{0,i}}=56\,{\rm{cm}}^{-3}$ per neutrino species $i=1,2,3$. Yet, in turn, the weakness of the neutrino interactions so far has spoilt all attempts to probe this $1.95$ K cosmic neutrino background (\CnuB), which is the analog of the $2.73$ K cosmic microwave background (CMB) of photons, in a laboratory setting~\cite{Hagmann:1999kf,Ringwald:2004te,Gelmini:2004hg,Ringwald:2005zf}. However, other cosmological measurements, such as the light element abundance, large scale structure (LSS) and the CMB anisotropies are sensitive to the presence of the \CnuB and therefore have provided us at least with indirect evidence for its existence (see e.g. Ref.~\cite{Hannestad:2004nb} for a review). 

Independently, Type Ia Supernova (SNIa) results (e.g.~\cite{Riess:2004nr}), supported by CMB~\cite{Spergel:2006hy} and LSS data (e.g. Refs.~\cite{Dodelson:2001ux,Szalay:2001rw}), strongly suggest the existence of an exotic, smooth energy component with negative pressure, known as dark energy, which drives the apparent accelerated expansion of our universe. Recently, Fardon, Nelson and Weiner~\cite{Fardon:2003eh} have shown that the relic neutrinos, which constitute the \CnuB, are promoted to a natural dark energy candidate if they interact through a new non-Standard-Model scalar force -- an idea which has great appeal. Neutrinos are the only Standard Model (SM) fermions without right-handed partners. Provided lepton number is violated, the active (left-handed) neutrinos are generally assumed to mix with a dark right-handed neutrino via the well-known seesaw mechanism~\cite{Gell-Mann,Yanagida,Minkowski,Mohapatra:1979ia}, thus opening a window to the dark sector. Therefore, it would not seem to be surprising if neutrinos, whose interactions and properties we know comparably little about, were sensitive to further forces mediated by dark particles. Moreover, the scale relevant for neutrino mass squared differences as determined from neutrino oscillation experiments, $\delta {\mnu}^2\sim (10^{-2}\,\,\eV)^2$, is of the order of the tiny scale associated with the dark energy, $(2\times 10^{-3}\,\,\eV)^4$.  

As a consequence of the new interaction in such a scenario, an intricate interplay links the dynamics of the relic neutrinos and the mediator of the dark force, a light scalar field called the acceleron. On the one hand, the neutrino masses $\mnui$ are generated by the vacuum expectation value $\A$ of the acceleron, $\mnui(\A)$. Correspondingly, the $\A$ dependence of the masses $\mnui(\A)$ is transmitted to the neutrino energy densities $\rho_{\nu_i}(\mnui(\A))$ since these are functions of $\mnui(\A)$. On the other hand, as a direct consequence, the neutrino energy densities $\rho_{\nu_i}(\mnui,\A)$ can stabilize the acceleron by contributing to its effective potential $\Veff(\A,\rhoi)$, which represents the total energy density of the coupled system. Moreover, cosmic expansion manifests itself in the dilution of the neutrino energy densities $\rhoi(z)\sim (1+z)^3$. Therefore, it crucially affects the effective acceleron potential $\Veff(\A,\rhoi(\mnui,z))$ by introducing a dependence on cosmic time, here parameterized in terms of the cosmic redshift $z$. For a homogeneous configuration, the equilibrium value of the acceleron instantaneously minimizes its effective potential $\Veff(\A,\rhoi(\mnui,z))$ and therefore also $\A(z)$ varies on cosmological time scales. Finally, in turn, since the neutrino masses $\mnui(\A)$ are sensitive to changes in $\A$, they are promoted to dynamical quantities depending on $z$, $\mnui(z)$, that is depending on cosmic time. To summarize, the variation of the neutrino masses represents a clear signature of the so-called Neutrino Dark Energy scenario.  

In a subsequent work Fardon, Nelson and Weiner~\cite{Fardon:2005wc} presented a supersymmetric Mass Varying Neutrino (MaVaN) model in which the origin of dark energy was attributed to the lightest neutrino $\nu_1$ and the size of the dark energy could be expressed in terms of neutrino mass parameters. By naturalness arguments the authors concluded that the lightest neutrino still has to be relativistic today, thereby preventing potential instabilities~\cite{Afshordi:2005ym,Takahashi:2006jt,Spitzer:2006hm} which could occur in highly non-relativistic theories of Neutrino Dark Energy. 

The rich phenomenology of the MaVaN scenario has been explored by many authors. The cosmological effects of varying neutrino masses have been studied in Refs.~\cite{Brookfield:2005td,Brookfield:2005bz} and were elaborated in the context of gamma ray bursts~\cite{Li:2004tq}. Apart from the time variation, the conjectured new scalar forces between neutrinos as well as the additional possibility of small acceleron couplings to matter lead to an environment dependence of the neutrino masses governed by the local neutrino and matter density~\cite{Fardon:2003eh,Kaplan:2004dq,Weiner:2005ac}. The consequences for neutrino oscillations in general were exploited in Refs.~\cite{Kaplan:2004dq,Zurek:2004vd} and in particular in the sun~\cite{Cirelli:2005sg,Gonzalez-Garcia:2005xu,Barger:2005mn}, in reactor experiments~\cite{Schwetz:2005fy,Gonzalez-Garcia:2005xu} as well as in long-baseline experiments~\cite{Gu:2005pq}. 

In light of the possible realization of Neutrino Dark Energy in nature, a (more) direct detection of the \CnuB should be thoroughly explored with special attention turned to possible new physics beyond the SM. By this means, a time evolution of neutrino masses could be revealed which would serve as a test of Neutrino Dark Energy. In addition, the general importance of a (more) direct evidence for the existence of the \CnuB lies in a confirmation of standard cosmology back to the freeze-out of the weak interactions and therefore thirteen orders of magnitudes before the time when photons where imprinted on the last scattering surface. 

An appealing opportunity to catch a glimpse of the \CnuB as it is today emerges from the possible existence of extremely high-energy cosmic neutrinos (\UHEnu's). Such \UHEnu's can annihilate with relic anti-neutrinos (and vice versa) into $Z$ bosons, if their energies coincide with the respective resonance energies $E_{0,i}^{\res}$ of the corresponding process $\nu\nubar\rightarrow Z$~\cite{Weiler:1982qy,Weiler:1983xx,Gondolo:1991rn,Roulet:1992pz,Yoshida:1996ie,Eberle:2004ua,Barenboim:2004di,D'Olivo:2005uh}. These energies, 
\be
\label{Eres0}
E_{0,i}^{\res}=\frac{M^2_Z}{2m_{\nu_{0,i}}}=4.2\times 10^{12}\,\,\left(\frac{\eV}{\mnui}\right) {\rm GeV}
\ee
in the rest system of the relic neutrinos, are entirely determined by the $Z$ boson mass $M_Z$ as well as the respective neutrino masses $\mnui$. An exceptional loss of transparency of the \CnuB for cosmic neutrinos results from the fact that the corresponding annihilation cross-section on resonance is enhanced by several orders of magnitude with respect to non-resonant scattering. As a consequence, the diffuse \UHEnu flux arriving at earth is expected to exhibit absorption dips whose locations in the spectrum are determined by the respective resonance energies of the annihilation processes. Provided that the dips can be resolved on earth, they could produce the most direct evidence for the existence of the \CnuB so far. Furthermore, as indicated by the resonance energies in Eq.~(\ref{Eres0}), the absorption features depend on the magnitude of the neutrino masses and could therefore reflect their possible variation with time. Moreover, they are sensitive to the flavor composition of the neutrino mass eigenstates as well as to various cosmological parameters. Accordingly, the possibility opens up to perform relic neutrino absorption spectroscopy as an independent means to probe neutrino physics and cosmology.

The existence of \UHEnu's is theoretically well motivated and is substantiated by numerous works on possible \UHEnu sources of astrophysical nature (bottom-up) (see e.g.~\cite{Torres:2004hk} for a review) or so-called top-down sources (see e.g. Ref.~\cite{Bhattacharjee:1998qc} for a review). In the latter case, \UHEnu's with energies well above $10^{11}$ GeV are assumed to be produced in the decomposition of topological defects (TD's) which originate from symmetry breaking phase transitions in the very early universe. 

\begin{figure}
\vspace* {0.0in}
\begin{center}
\includegraphics*[bbllx=12pt,bblly=208pt,bburx=577pt,bbury=611pt,height=9.5cm,width=14.5cm]{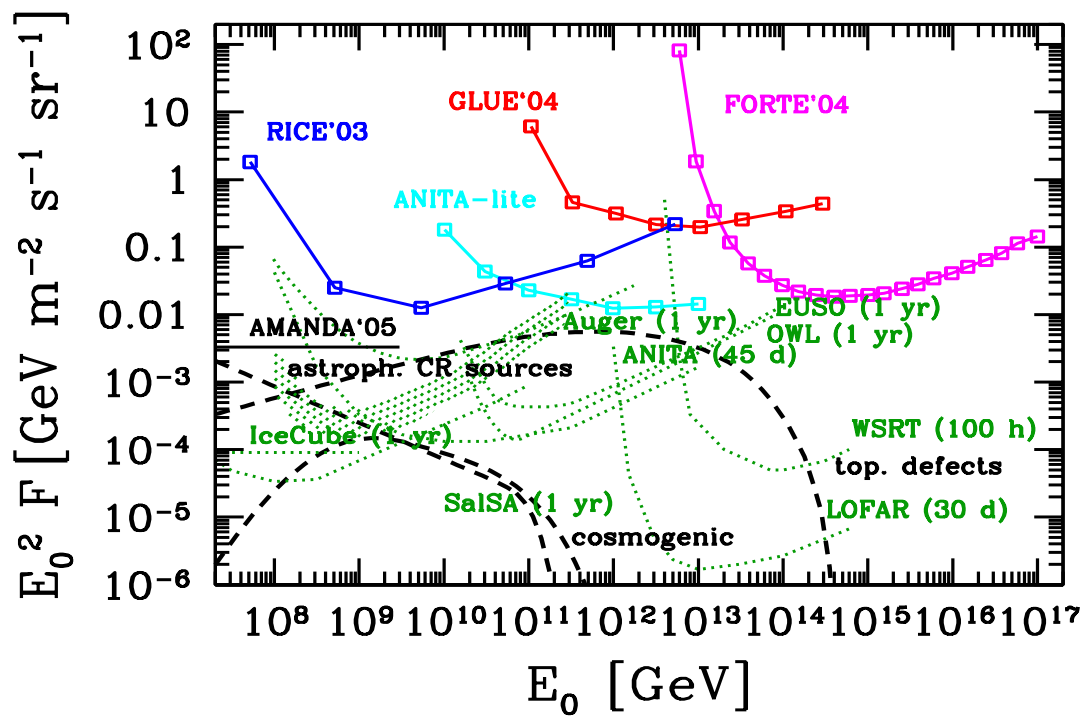}
\caption[]{Current status and next decade prospects for EHEC$\nu$ physics, 
expressed in terms of diffuse neutrino fluxes per flavor, 
$F=F_{\nu_\alpha}+F_{\bar\nu_\alpha}$, $\alpha =e,\mu,\tau$.
The upper limits from  AMANDA~\cite{Ackermann:2005sb}, see also Ref.~\cite{Halzen:2006ic}, ANITA-lite~\cite{Barwick:2005hn}, 
FORTE~\cite{Lehtinen:2003xv}, GLUE~\cite{Gorham:2003da}, and RICE~\cite{Kravchenko:2003tc} are plotted.
Also shown are projected sensitivities of 
ANITA\cite{Barwick:2005hn}, EUSO~\cite{Bottai:2003ep}, IceCube~\cite{Ahrens:2002dv}, 
LOFAR~\cite{Scholten:2005pp}, OWL~\cite{Stecker:2004wt}, 
the Pierre Auger Observatory in $\nu_e$, $\nu_\mu$ modes and in $\nu_\tau$ mode 
(bottom swath)\cite{Bertou:2001vm},  
SalSA~\cite{Gorham:2001wr}, and WSRT\cite{Scholten:2005pp}, corresponding to 
one event per energy decade and indicated duration.   
Also shown are predictions from astrophysical Cosmic Ray (CR) sources\cite{Ahlers:2005sn}, 
from inelastic interactions of CR's with the cosmic microwave background (CMB) photons 
(cosmogenic neutrinos)\cite{Ahlers:2005sn,Fodor:2003ph}, 
and from topological defects\cite{fodor:tbp}.}
\end{center}
\label{fig1}
\end{figure}  

Depending on the underlying \UHEnu sources the \UHEnu fluxes could be close to the current observational bounds set by existing \UHEnu observatories such as AMANDA~\cite{Ackermann:2005sb} (see also Ref.~\cite{Halzen:2006ic}), ANITA-lite\cite{Barwick:2005hn}, BAIKAL~\cite{Wischnewski:2005rr}, FORTE~\cite{Lehtinen:2003xv}, GLUE~\cite{Gorham:2003da} and RICE~\cite{Kravchenko:2003tc} which cover an energy range of $10^7\,\,{\rm GeV}<E_0<10^{17}\,\,{\rm GeV}$ (cf. Fig.~\ref{fig1}). Promisingly, the sensitivity in this energy range will be improved by orders of magnitude (cf. Fig.~\ref{fig1}) by larger \UHEnu detectors such as ANITA, EUSO~\cite{Bottai:2003ep}, IceCube~\cite{Ahrens:2002dv}, LOFAR~\cite{Scholten:2005pp}, OWL~\cite{Stecker:2004wt}, SalSA~\cite{Gorham:2001wr} and WRST~\cite{Scholten:2005pp} which are planned to start operating within the next decade (cf. Fig.~\ref{fig1}). Accordingly, the prospects of confirming the existence of the \CnuB by tracking its interaction with \UHEnu's have substantially improved since the original proposal in 1982~\cite{Weiler:1982qy}. Moreover, in the likely case of appreciable event samples the valuable information encoded in the absorption features of the \UHEnu spectra could be revealed within the next decade (cf. Fig.~\ref{fig1}), rendering the theoretical exploration of relic neutrino absorption spectroscopy a timely enterprise.
 
Note that the scenario introduced above has also attracted attention for another reason than the possible detection of the \CnuB -- namely for the controversial possibility of solving the so-called GZK-puzzle to be discussed briefly in the following. Beyond the Greisen-Zatsepin-Kuzmin (GZK) energy, $E_{\rm GZK}=4\times 10^{10}$ GeV, ultra-high energy nucleons rapidly lose energy due to the effective interaction with CMB photons (predominantly through resonant photo-pion production)~\cite{Greisen:1966jv,Zatsepin:1966jv}. In the so-called $Z$-burst scenario, the observed mysterious cosmic rays above $E_{\rm GZK}$ were associated with the secondary cosmic ray particles produced in the decays of $Z$ bosons. The latter were assumed to originate from the neutrino annihilation process outlined above~\cite{Fargion:1997ft,Weiler:1997sh,Yoshida:1998it,Fodor:2001qy,Fodor:2002hy,Gelmini:2004zb}.

However, recently, ANITA-lite~\cite{Barwick:2005hn} appears to have entirely excluded the $Z$-burst explanation for the GZK-puzzle at a level required to account for the observed fluxes of the highest energy cosmic rays. We would like to stress, that this only means that the GZK-puzzle stays unsolved (if there is any) . Moreover, this neither restricts the possible success of producing evidence for the \CnuB by means of detecting absorption dips in the \UHEnu spectra nor does it spoil the possibility of gaining valuable information from performing relic neutrino absorption spectroscopy.

The goal of this paper is to carefully work out the characteristic differences in the \UHEnu absorption features which result from treating the neutrino masses as time varying dynamical quantities in comparison to constants. In our analysis, we incorporate the full thermal background effects on the absorption process whose impact grows for smaller neutrino masses~\cite{Barenboim:2004di,D'Olivo:2005uh}. This means, that in general relic neutrinos cannot be assumed to be at rest. Instead, they have to be treated as moving targets with a momentum distribution, if their mean momenta turn out to be of the order of the relic neutrino masses.

We illustrate our results for the diffuse neutrino fluxes to be observed at earth firstly by considering astrophysical \UHEnu sources. Secondly, we calculate the neutrino spectrum (both for varying and constant neutrino masses) expected from the decomposition of a topological defect using the appropriate fragmentation functions and including the full thermal background effects. By this means, for a given \UHEnu injection spectrum, we present all technical tools to interpret \UHEnu absorption dips as soon as they are observed at earth. Thereby the possibility opens up to test the \CnuB and its interpretation as source of Neutrino Dark Energy. 

The paper is organized as follows. In Sec.~1 we discuss the MaVaN scenario and focus on a determination of the time dependence of the neutrino masses. Sec.~2 reviews the treatment of the absorption of an \UHEnu by the \CnuB in terms of the damping factor comprising the thermal background effects. Furthermore, in order to make contact to treatments in the literature~\cite{Weiler:1982qy,Weiler:1983xx,Roulet:1992pz,Eberle:2004ua,Barenboim:2004di,D'Olivo:2005uh}, we include in the discussion common approximations~\cite{Roulet:1992pz,Eberle:2004ua,Barenboim:2004di} which neglect part or all of the dependence of the damping on the relic neutrino momenta. Moreover, we extend the complete analysis to incorporate a possible variation of the neutrino masses with time. In Sec.~3 we present and compare our results for the survival probabilities of \UHEnu's with varying and constant masses which encode the physical information on all possible annihilation processes on their way from their source to us, again taking into account the thermal motion of the relic neutrinos. In order to gain more physical insight, in addition, we disentangle the characteristic features of the absorption dips caused by the mass variation by switching off all thermal background effects. Sec.~4 illustrates the discovery potential of neutrino observatories for the \CnuB and gives an outlook for the testability of the MaVaN scenario. Therefore, both for astrophysical sources and for a topological defect scenario, we calculate the expected observable \UHEnu flux arriving at earth which results from folding the survival probabilities with the corresponding \UHEnu source emissivity distribution. In the latter case, we perform the full state-of-the-art calculation with the help of fragmentation functions and by the inclusion of all thermal background effects. In Sec.~5 we summarize our results and conclude. 

\section{Mass Varying Neutrinos (MaVaNs)\label{sec:MaVaNs}}%

In Ref.~\cite{Fardon:2003eh} a new non-Standard Model interaction between neutrinos and a light `dark' scalar field, the so-called acceleron, was introduced. In essence, it serves as possible origin of the apparent accelerated expansion of the universe and promotes the \CnuB to a natural dark energy candidate. Furthermore, as a very interesting and intriguing secondary effect, it causes a time evolution of neutrino masses. 
 
A follow up publication~\cite{Fardon:2005wc} takes care of a possible stability problem of the model~\cite{Afshordi:2005ym,Takahashi:2006jt,Spitzer:2006hm} and furnishes a viable model of the whole scenario.

Largely following Refs.~\cite{Fardon:2003eh,Fardon:2005wc}, in this section we discuss the details of the complex interplay between the acceleron and the neutrinos that arises from a Yukawa coupling between them. Thereby, we will mainly focus on the determination of the resulting time variation of neutrino masses to be implemented later on in our analysis on relic neutrino absorption. For the latter it will turn out that the results are largely independent of the details of the model, since only a few generic features of the setting enter the investigation. 

The new acceleron-neutrino interaction has a twofold effect. On the one hand, as a direct consequence, the neutrino masses $\mnui$ are generated from the vacuum expectation value (VEV) $\A$ of the acceleron and become functions of $\A$, $\mnui(\A)$, $i=1,2,3$. On the other hand, the dependence of $\mnui$ on $\A$ turns the neutrino energy densities $\rhoi$ into implicit functions of $\A$, since the energy densities $\rhoi\,(\mnui(\A))$ depend on the masses $\mnui(\A)$, $i=1,2,3$. In this way, the energy density contained in a homogeneous background of neutrinos can stabilize the acceleron by contributing to its effective potential $\Veff(\A)$. In other words, the dependence of the free energy on the value of $\A$ gets a contribution from the rest energy in neutrinos in addition to the pure scalar potential $V_0(\A)$. The total energy density of the system parameterized by the effective acceleron potential $\Veff(\A)$ takes the following form,
\be
\Veff(\A)=\sum\limits^3_{i=1}\rhoi(\mnui(\A),z)+V_0(\A).\label{Veff}
\ee
This is to be contrasted with the situation in empty space: if $V_0(\cal{\A})$ is a `run-away potential', the acceleron does not possess a stable vacuum state but rolls to its state of lowest energy given by the minimum of its pure potential $V_0(\A)$. 

Taking now the expansion of the universe into account, the dilution of the neutrino energy densities $\rhoi(z)\sim (1+z)^3$ introduces a time dependence (here parameterized in terms of the cosmic redshift $z$) into the effective acceleron potential $\Veff$. Consequently, in the adiabatic limit \footnote{Under the assumption that the curvature scale of the potential is much larger than the expansion rate, $\partial ^2 \Veff(\A)/\partial {\A}^2=m_{\A}^2\gg H^2$, the adiabatic solution to the equations of motion applies. In this case for $|\A|<M_{Pl}\simeq 3\times 10^{18}$ GeV the effects of the kinetic energy terms can be safely ignored~\cite{Fardon:2003eh}.}, the equilibrium value $\A$ of the acceleron has to vary with time in order to instantaneously minimize its effective potential $\Veff(\A)$\footnote{Since therefore in the presence of the relic neutrinos the acceleron possesses a stable (time dependent) vacuum state, in the literature both the acceleron and its VEV are referred to as $\A$.}. Finally, as the neutrino masses $\mnui(\A)$ are directly affected by changes in the $\A$ condensate, they are promoted to dynamical quantities $\mnui(z)$ depending on cosmic time.

Note that Eq.~(\ref{Veff}) takes the neutrino energy density $\rhoi$ to be spatially constant. To justify this assumption, the $\A$ condensate is not allowed to vary significantly on distances of the order of the inter-neutrino spacing $r$ of the relic neutrinos, with currently $1/r \simeq 336^{1/3} {\rm cm}^{-1}$, where we have assumed a neutrino and anti-neutrino number density of $n_{\nu_{0,i}}=n_{\nubar_{0,i}}\simeq 56\,{\rm cm}^{-3}$ per species $i=1,2,3$. Consequently, remembering that the range of the force mediated by a scalar field is equal to its inverse mass, one arrives at an upper bound on the $\A$ mass $m_{\A}$ given by $m_{\A}<1/r\sim O(10^{-4}\,\eV)$ at the present time. 

Let us now determine the time evolution of the physical neutrino masses $\mnui(z)$. Since the neutrino masses arise from the instantaneous equilibrium value $\A$, we have to analyze the minimum of the total energy density $\Veff(\A)$. Assuming $\frac{\partial \mnui(\A)}{\partial\A}$ to be non-vanishing, one arrives at,
\be
\label{minimum}
\frac{\partial \Veff(\A)}{\partial\A}=\left. \sum\limits^3_{i=1}\frac{\partial \rhoi(\mnui,z)}{\partial  \mnui}\right|_{\mnui=\mnui(\A)}\frac{\partial \mnui(\A)}{\partial\A}+\frac{\partial V_0(\A)}{\partial\A}=0,
\ee
where~\cite{Peccei:2004sz}
\bea
\rhoi(\mnui,z)&=&\frac{\T0^4}{\pi^2}\,(1+z)^4 \int\limits_0^{\infty}\frac{dy\, y^2 \sqrt{y^2+x^2_i}}{e^y+1}\hspace{4ex}\mbox{and}\\
x_i&=&\frac{\mnui}{\T0\,(1+z)}, 
\eea
with $\T0=1.69\times 10^{-4}$ eV denoting the neutrino temperature today. Note that the condition for the minimal energy density leads to a dependence of the neutrino masses on the neutrino energy densities which evolve with $z$ on cosmological time scales.

The smallness of the active neutrino masses $\mnui$ can be explained by letting them only indirectly feel the acceleron via the seesaw mechanism~\cite{Gell-Mann,Yanagida,Minkowski,Mohapatra:1979ia}. Therefore, following Refs.~\cite{Fardon:2003eh,Fardon:2005wc}, we introduce three `right-handed' or `sterile' neutrinos $N_i$ with no Standard Model charges, whose masses $M_{N_i}$ are constructed to vary with $\A$ due to a direct Yukawa interaction. In the seesaw mechanism the active neutrino masses $\mnui$ are functions of the sterile neutrino masses $M_{N_i}(\A)$. Consequently, the $\A$ dependence of the $M_{N_i}(\A)$ is transmitted to the active neutrino masses $\mnui(\A)$ and causes them to change accordingly. Let us consider the interaction~\cite{Fardon:2003eh,Fardon:2005wc}:

\be
{\cal{L}}\supset m_{D_{ij}}N_i\nu_{l_j}+\kappa_{ij} N_i N_j\A+h.c.+V_0(\A).
\ee
where $i,j=1,2,3$ are the family-number indices and $\nu_{l_i}$ correspond to the left-handed active neutrinos. Furthermore, $\kappa\A$ corresponds to the $\A$ dependent mass matrix of the sterile neutrinos and $m_{D}$ is the Dirac type matrix (originating from electroweak symmetry breaking). Assuming the eigenvalues of $\kappa\A$ to be much larger than the eigenvalues of $m_D$ one can integrate out the sterile neutrinos $N_i$, arriving at the following effective low energy Lagrangian~\cite{Fardon:2003eh,Fardon:2005wc},
\bea
{\cal{L}}&\supset&M_{ij}(\A)\,\nu_{l_i}\nu_{l_j}+h.c.+V_0(\A),\,\,\mbox{where}\\
M_{ij}(\A)&=& \frac{(m^T_D\kappa^{-1}m_D)_{ij}}{\A}\label{M}
\eea
represents the mass matrix of the active neutrinos.

In order to solve Eq.~(\ref{minimum}) for $\mnui(z)$ and to do MaVaN phenomenology the fundamental scalar potential $V_0(\A)$ has to be specified in an appropriate way. Namely, the coupled neutrino acceleron fluid has to act as a form of dark energy which is stable against growth of inhomogeneities~\cite{Afshordi:2005ym} and, as suggested by observations, must redshift with an equation of state $\omega \sim -1$ today.
 
An appealing possibility arises in the framework of so-called hybrid models~\cite{Linde:1993cn}. Those models were introduced to explain accelerated expansion in the context of inflation. In essence, two light scalar fields interact in such a way that one of them stabilizes the other one in a metastable minimum. The energy density stored in the potential associated with the false minimum can drive accelerated expansion. 

It turns out, that a straightforward supersymmetrization~\cite{Fardon:2005wc} of the MaVaN model naturally sets the stage to apply the idea of the acceleration mechanism to dark energy. Identifying the acceleron with the former of the two light scalar fields, the hybrid model provides a microscopic origin for a quadratic acceleron potential $V_0(\A)\sim \A^2$. The role of the residual light scalar field coupled to the acceleron is attributed to the scalar partner $\N$ of a sterile neutrino naturally present in a supersymmetric theory. The acceleron is stabilized by the presence of the fermionic neutrino background which drives its VEV to larger values. Accordingly, acceleration lasts as long as the VEV of the acceleron is sufficiently high to keep the sterile sneutrino $\N$ in a false metastable minimum. So long as the energy density stored in the $\A$ condensate is sufficiently small, the combined scalar potential $V(\N,\A)$ will appear as dark energy redshifting with an equation of state $\omega\sim -1$~\cite{Fardon:2005wc}. Consequently, the neutrino dark energy density $\Omega_X\sim \rm{const.}$ cosmologically behaves very much like a cosmological constant.  

According to Ref.~\cite{Fardon:2005wc} naturalness arguments require $\N\equiv{\cal{N}}_1$, assigning the lightest sterile sneutrino ${\cal{N}}_1$ to be responsible for dark energy. Furthermore, one can conclude that ${\cal{N}}_1$ has to be at least moderately relativistic today ($m_{\nu_{0,1}}\lwig \T0$). Accordingly, in this supersymmetric MaVaN model probable instabilities~\cite{Afshordi:2005ym,Takahashi:2006jt,Spitzer:2006hm} of highly non-relativistic MaVaN theories do not occur. 

In the past, the heavier two sterile sneutrinos ${\cal{N}}_2,{\cal{N}}_3$ of the theory were stabilized by the acceleron like the lightest ${\cal{N}}_1$. However, by today they are assumed to have reached their state of lowest energy having acquired vacuum expectation values.\footnote{We refer to the mechanism proposed in~\cite{Fardon:2005wc} to evade large ${\cal{N}}_{2,3}$ VEV contributions to the $\A$ mass conflicting with the upper mass bound set by the current inter-neutrino spacing $O(10^{-4})$ eV.} 

The relevant contribution~\cite{Fardon:2005wc} to the superpotential is given in terms of couplings of the superfield containing the acceleron $\A$ to two superfields with generation indices $i$ and $j$, with $i,j=1,2,3$, which comprise the sterile neutrinos $N_i$,$N_j$ and their respective scalar partners ${\cal {N}}_i$ and ${\cal {N}}_j$. The coupling constant matrix has elements $\kappa_{ij}$. This superfield interaction provides the necessary couplings mentioned above, namely of the scalar acceleron to the sterile sneutrino fields as well as the scalar acceleron Yukawa coupling to the sterile neutrinos in terms of $\kappa_{ij}$. In Ref.~\cite{Fardon:2005wc} the one loop radiative corrections were estimated and taken to give a lower bound on the natural size for the magnitudes of the soft susy breaking masses squared of the $\A$ and the ${\cal{N}}_i$.

Given the exploratory nature of this investigation and following largely Ref.~\cite{Fardon:2005wc} it is reasonable to exploit the rough proportionality $\delta m^2_{\N_i}\sim - m^2_{D_i}$ to get an estimate of the $\A$ mass squared $m^2_{\A}$; here $\delta m^2_{{\cal {N}}_i}$ represent the one loop radiative corrections to the mass of a sterile sneutrino ${\cal {N}}_i$ and $m_{D_i}$ label the respective eigenvalues of the Dirac type matrix $m_D$. For simplicity, we assume no off-diagonal elements for the coupling constant matrix $\kappa$ and denote the diagonal matrix elements by $\kappa_i$. Finally, one arrives at an estimate for the acceleron mass squared  $m^2_{\A}$~\cite{Fardon:2005wc},
\be
m^2_{\A}\sim \sum\limits^3_{i=1} \kappa^2_i m^2_{D_i},
\ee
such that the quadratic acceleron potential can be expressed in terms of neutrino mass parameters according to
\be
\label{V0}
V_0(\A)\sim m^2_{\A}{\A}^2=\sum\limits^3_{i=1} \kappa^2_i m^2_{D_i}{\A}^2.
\ee

Now we are in a position to determine the respective mass-redshift relations $\mnui(z)$ of the active MaVaNs whose mass squared differences today have to be compatible with oscillation experiments. Taking the acceleron Yukawa matrix $\kappa$ as well as the Dirac type mass matrix $m_D$ in Eq.~(\ref{M}) to be diagonal, one arrives at the approximate seesaw formula for the physical neutrino masses:
\be
\label{seesaw}
\mnui(\A)=\frac{m^2_{D_i}}{\kappa_i \A},\,\,\mbox{where}\,\,i=1,2,3.
\ee
Accordingly, the instantaneous minimum of $\Veff$ in Eq.~(\ref{minimum}) is determined by
\be
\frac{\partial \Veff(\A)}{\partial\A}=\left. \sum\limits^3_{i=1}\frac{\partial \rhoi(\mnui,z)}{\partial  \mnui}\right|_{\mnui=\frac{m^2_{D_i}}{\kappa_i \A}}\left(\frac{-m^2_{D_i}}{\kappa_i \A^2}\right)+2\sum\limits^3_{i=1}\kappa^2_i m^2_{D_i}\A=0.
\ee
Since this equation has to hold for all $z$, the acceleron VEV becomes a function of $z$. As a direct consequence, it generates redshift dependent neutrino masses $\mnui(z)$,
\be
\mnui(z)=\frac{m^2_{D_i}}{\kappa_i \A(z)}\,\,\mbox{with}\,\,\mnui(0)=m_{\nu_{i_0}}=\frac{m^2_{D_i}}{\kappa_i \A(0)},
\ee
which implies
\be
\label{massA}
\mnui(z)=m_{\nu_{0,i}}\frac{\A(0)}{\A(z)}.
\ee
Note that in general a MaVaN mass with subscript $0$ has to be identified with the present day neutrino mass. Consequently, the $m_{\nu_{0,i}}$ have to be consistent with the mass squared differences measured in neutrino oscillation experiments. 

Accordingly, inserting Eq.~(\ref{massA}) and Eq.~(\ref{V0}) into Eq.~(\ref{minimum}) yields, 
\be
\label{massevolution}
\left. \sum\limits_{i=1}^{3}\, m_{\nu_{0,i}} \frac{\partial \rhoi(\mnui,z)}{\partial \mnui}\right|_{\mnui=m_{\nu_{0,i}}\frac{\A(0)}{\A(z)}}-\left(\frac{\A(z)}{\A(0)}\right)^3 2\sum\limits_{i=1}^{3}\,\frac{m^6_{D_i}}{m^2_{\nu_{0,i}}}=0.
\ee 
Evaluating Eq.~(\ref{massevolution}) at $z=0$, 
\be\label{m}
\hspace{0cm}2\sum\limits_{i=1}^{3}\frac{m^6_{D_i}}{m^2_{\nu_{0,i}}}=\left.\sum\limits_{i=1}^{3}m_{\nu_{0,i}}\frac{\partial \rhoi(\mnu0,i)}{\partial \mnui}\right|_{\mnui=m_{\nu_{0,i}}},
\ee 
allows to eliminate $2\sum\limits_{i=1}^{3}\,\frac{m^6_{D_i}}{m^2_{\nu_{0,i}}}$ from Eq.~(\ref{massevolution}) leading to
\be
\label{finalmassevolution}
\hspace{-2cm}\left.\sum\limits_{i=1}^{3}\, m_{\nu_{0,i}}\,\Big(\frac{\partial \rhoi(\mnui,z)}{\partial \mnui}\right|_{\mnui=m_{\nu_{0,i}}\frac{\A(0)}{\A(z)}}-\left.\left(\frac{\A(z)}{\A(0)}\right)^3\,\frac{\partial \rhoi(\mnui,0)}{\partial \mnui}\right|_{\mnui=m_{\nu_{0,i}}}\Big)=0.
\ee
Finally, the solution for $\left(\frac{\A(z)}{\A(0)}\right)^3$, which can only be determined numerically, fixes the neutrino mass evolution according to Eq.~(\ref{massA}) in terms of the present day neutrino masses $m_{\nu_{0,i}}$. 

However, the mass behavior in the low as well as in the high redshift regime can be approximated analytically by using the respective limits for $\frac{\partial \rhoi}{\partial \mnui}$. As mentioned before, in the supersymmetric MaVaN model~\cite{Fardon:2005wc} the lightest neutrino is at least moderately relativistic today such that its mass has to be very small, $m_{\nu_{0,1}}\lwig \T0=1.69\times 10^{-4}$ eV. Furthermore, it can be deduced from the mass squared differences measured at neutrino oscillation experiments (see~\cite{Strumia:2006db} for a recent review) that the heavier two neutrinos are non-relativistic today ($m_{\nu_{0,i}}\gg \T0$ for $i=2,3$).

Accordingly, in the low redshift regime it is a good approximation to neglect the contribution of the lightest neutrino species to $\Veff$ and solely employ the non-relativistic limit of $\frac{\partial \rhoi}{\partial \mnui}$ with $i=2,3$ and $x_i=\frac{\mnui}{\T0 (1+z)}\gg 1$,
\be
\label{NR}
\frac{\partial \rhoi(\mnui,z)}{\partial \mnui}\approx \frac{\T0^3}{\pi^2} (1+z)^3 \int\limits_{0}^{\infty}\frac{y^2}{e^y+1} dy.
\ee
Accordingly, in the low redshift regime Eq.~(\ref{finalmassevolution}) is solved by
\bea
\A_{\rm low}(z)&=&\A(0)\,(1+z)\\
\rightarrow m_{\nu_{i,\rm low}}(z)&=& m_{\nu_{0,i}}(1+z)^{-1},\,\,\mbox{i=1,2,3}, 
\label{LR}
\eea
where Eq.~(\ref{massA}) was used.

Once in the past all neutrinos were relativistic. In this regime, $x_i\ll 1$, $\frac{\partial \rhoi}{\partial \mnui}$ can be approximated by,
\be
\label{Rel}
\frac{\partial \rhoi(\mnui,z)}{\partial \mnui}\approx \frac{\T0^2}{\pi^2} (1+z)^2 \mnui \int\limits_{0}^{\infty}\frac{y}{e^y+1} dy.
\ee
By taking the appropriate approximations Eq.~(\ref{NR}) and Eq.~(\ref{Rel}) for the two terms in Eq.~(\ref{finalmassevolution}), $\Veff$ is minimized for,
\bea
\A_{\rm high}(z)&\propto&(1+z)^{1/2}\\
\rightarrow m_{\nu_{i,\rm high}}(z)&\propto& (1+z)^{-1/2},\,\,\mbox{i=1,2,3},
\label{HR}
\eea
where the factor of proportionality is a function of the present day neutrino masses and the integrals in Eq.~(\ref{NR}) and Eq.~(\ref{Rel}).

In our analysis on cosmic neutrino absorption later on, these approximations will help towards a better understanding of the numerical calculations since the corresponding results agree very well in the respective regimes. 

\begin{figure}
\vspace* {0.0in}
\begin{center}
\includegraphics*[bbllx=20pt,bblly=150pt,bburx=580pt,bbury=678pt,height=10.3cm,width=10.9cm]{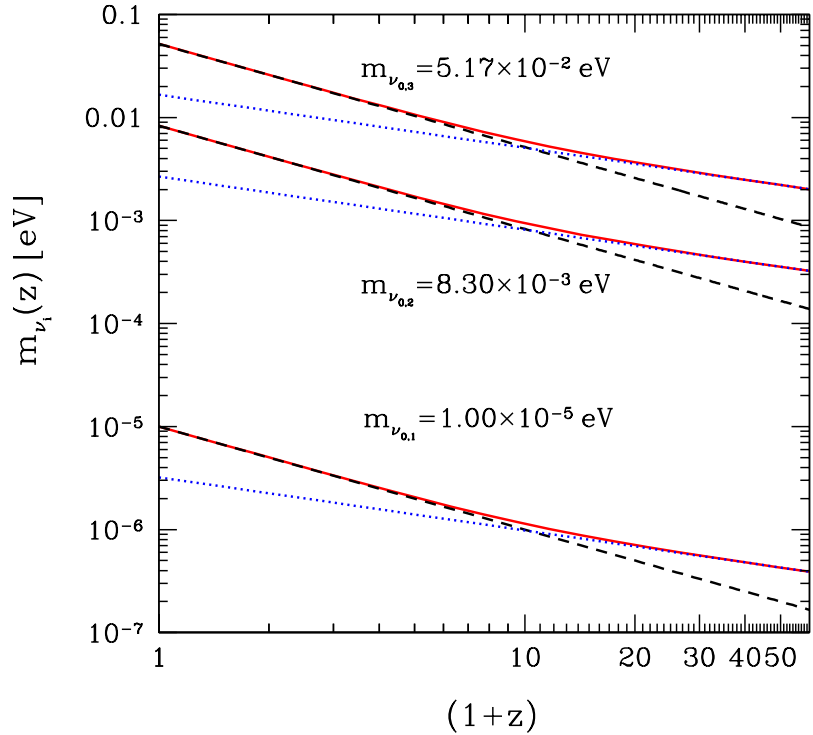}
\caption[]{In this log-log plot the exact mass-redshift relations $\mnui(z)$ are plotted as solid lines. In the low as well as in the high redshift regime they are well approximated by simple power laws, $\mnui(z)\propto (1+z)^{-1}$ and $\mnui(z)\propto (1+z)^{-1/2}$, respectively (dashed and dotted lines). The lightest MaVaN is assumed to have a mass of $m_{\nu_{0,1}}=10^{-5}\,\eV$ today. Consequently, for a normal neutrino mass hierarchy, solar and atmospheric mass splittings fix the present time neutrino masses of the heavier neutrinos to be $m_{\nu_{0,2}}=8.3\times 10^{-3}\,\eV$ and $m_{\nu_{0,3}}=5.17\times 10^{-2}\,\eV$.}
\label{fig2}
\end{center}
\end{figure}

As becomes apparent from the approximations in Eq.~(\ref{LR}) and Eq.~(\ref{HR}) the MaVaN masses $\mnui(z)$ are decreasing functions of redshift. Correspondingly, MaVaNs can be regarded as practically massless in the past whereas in the present epoch they have reached their maximal mass values $m_{\nu_{0,i}}$. In the following we will assume a normal neutrino mass hierarchy and take the mass of the lightest neutrino to be 
\be
\label{m1}
m_{\nu_{0,1}}=10^{-5}\,\eV, 
\ee
such that it is still relativistic today (i.e. $m_{\nu_{0,1}}\le \T0$). According to the solar and atmospheric neutrino mass splittings the corresponding present time masses of the heavier non-relativistic neutrinos are fixed to be (see e.g.~\cite{Strumia:2006db}) 
\bea
m_{\nu_{0,2}}=8.30\times 10^{-3}\,\eV\gg \T0,\\
m_{\nu_{0,3}}=5.17\times10^{-2}\,\eV \gg \T0.
\label{m3} 
\eea
Figure~\ref{fig2} shows that the exact mass-redshift relations $m_{\nu_i}(z)$, $i=1,2,3$, as numerically determined from Eq.~(\ref{finalmassevolution}) in combination with Eq.~(\ref{massA}), are very well approximated in the low as well as in the high redshift regime by simple power laws stated in Eq.~(\ref{LR}) and in Eq.~(\ref{HR}), respectively. These results have to be contrasted with the standard time-independent neutrino masses $m_{\nu_{0,i}}$ for $i=1,2,3$. 

Let us summarize the essential features of the presented viable MaVaN model which will enter the analysis on relic neutrino absorption later on. Firstly, the lightest neutrino is required to be still moderately relativistic today ($m_{\nu_{0,1}}\le \T0=1.69\times 10^{-4}\,\eV$) and therefore fixes the neutrino mass scale to be low. Secondly, the specific mass-redshift evolution $\mnui(z)$ is determined by the model dependent quadratic form of the potential $V_0\sim \A^2$ in Eq.~(\ref{V0}) which enters Eq.~(\ref{m}). The masses behave like $\propto (1+z)^{-1}$ and $\propto (1+z)^{-1/2}$ in the low and in the high redshift regime, respectively (cf. Eq.~(\ref{LR}) and Eq.~(\ref{HR})). However, the generically important feature (of any standard MaVaN model) is that the MaVaN masses $\mnui(z)$ are decreasing functions of redshift which is easily achieved within the framework of the seesaw mechanism (see above). As will turn out in our investigation later on, for decreasing $\mnui(z)$ the results do not strongly dependent on the actual functional dependence $\mnui(z)$. Note that the characteristic differences in the absorption features of MaVaNs with respect to constant mass neutrinos (cf. Sec.\ref{sec:SurvivalProb} and Sec.\ref{sec:AbsorptionDips}) are independent of the supersymmetrization of the MaVaN scenario.

\section{Damping of Extremely High-Energy Cosmic Neutrinos Traveling Through the Cosmic Neutrino Background \label{sec: Damping}}%

Due to the feebleness of the weak interaction cosmic neutrinos can propagate cosmological distances through the cosmic microwave and neutrino background (\CMB and \CnuB) without a significant probability of interacting. 

An interesting exception arises under the assumption that \UHEnu's with energies of order $E_{0,i}^{\res}=M_Z^2/2 \mnui=4.2\times 10^{14}\,{\rm GeV}\,(0.01 {\rm eV}/\mnui )$ in the rest system of the target $\nubar$ exist, where $M_Z$ is the $Z$ mass and $\mnui$ the respective neutrino mass. The $Z$ resonance in the $s$ channel for the process $\nu\nubar\rightarrow X$, characterized by the energy $E_{0,i}^{\res}$, enhances the cross-section for the annihilation of such an \UHEnu on a big-bang relic anti-neutrino (and vice versa) by orders of magnitude. As a consequence, the corresponding interaction probability significantly increases with respect to non-resonant scattering. Accordingly, the annihilation processes would reduce the survival probability of \UHEnu's traveling through the \CnuB to us and could therefore be detectable as absorption dips in the \UHEnu spectra. 

In this section we consider \UHEnu's (on whose plausible sources we will comment in Sec.~\ref{sec:AbsorptionDips}) propagating through a thermal bath of relic neutrinos in the expanding universe. Following Ref.~\cite{D'Olivo:2005uh}, we discuss the corresponding damping rate of the \UHEnu's which governs their survival probability (cf. Sec.~\ref{sec:SurvivalProb}). Furthermore, we will summarize common approximations for the damping which result from averaging over the neutrino momenta~\cite{Barenboim:2004di} or from completely neglecting the relic neutrino motion~\cite{Weiler:1982qy,Weiler:1983xx,Roulet:1992pz,Eberle:2004ua}. 

The investigation applies to both constant mass neutrinos as well as to MaVaNs. In the latter case, the neutrino masses $\mnui$ are not constant but -- as discussed in Sec.~\ref{sec:MaVaNs} -- complicated functions of the neutrino densities and thus functions of $z$ (cf. Eq.~(\ref{finalmassevolution}) in combination with Eq.~(\ref{massA}) in Sec.~\ref{sec:MaVaNs}).  

For simplicity, throughout this section we drop the indices $i=1,2,3$ labeling the mass, energy or momentum of the neutrino mass eigenstates, since the discussion applies to all three neutrinos likewise ($\mnu$ corresponds to $\mnui$, $E$ to $E_{i}$ etc.).  

The crucial quantity which describes the attenuation of an ultra-relativistic \UHEnu neutrino traversing the \CnuB is the damping rate $\gamma_{\nu\nubar}$. It can be expressed in the following form\footnote{The formula for the cross-section as well as the damping rate in Eq.~(\ref{eq:gammasigma}) apply likewise if both of the interacting particles are replaced by their anti-particles.},
  \begin{equation}
\label{eq:gammasigma} 
\gamma_{\nu\nubar}(E) = \int_0^\infty \frac{dP}{2\pi^2}
\ P^2 \ f_\nubar (P,\Tnu) \ \sigma_{\nu\nubar} (P,E),
\end{equation}
and is governed by the $Z$-formation cross-section $\sigma_{\nu\nubar} (P,E)$ weighted by the momentum distribution $f_\nubar (P,\Tnu)$ of the relic anti-neutrinos, both to be discussed in the following. Here, $P$ denotes the modulus of the relic anti-neutrino momentum, whereas $E$ represents the \UHEnu energy and $\Tnu$ is the \CnuB temperature. Note that in the expanding universe these quantities, as well as the relic neutrino energy $E_p=\sqrt{P^2+\mnu^2}$, are subject to cosmic redshift. They can be expressed in terms of their present day values, labeled by a subscript $0$, in the following way 
\be
\hspace{-1cm}P=P_0(1+z),\,\, E=E_0(1+z),\,\, \Tnu=\T0(1+z)\,\,\mbox{and}\,\, E_p=E_{p_0}(1+z), 
\ee 
where we take the \CnuB temperature at present to be $\T0=1.69 \times 10^{-4}$ eV.

Let us first consider the momentum distribution of the relic (anti-)neutrinos. Since they were ultra-relativistic at the time of decoupling their distribution assumes the relativistic Fermi-Dirac form~\cite{Pastor:2005av} both for constant as well as varying neutrino masses. Neglecting the chemical potential of the relic (anti)-neutrinos it is given by
\be
\label{eq:FDdistri}
f_{\nubar}(P,\Tnu)=f_{\nu}(P,\Tnu)=\frac{1}{e^{P/\Tnu}+1}.
\ee
Note that $P/\Tnu=P_d(1+z)/T_{\nu_d}(1+z)$ is a non-redshifting quantity where the subscript $d$ denotes the parameter values at decoupling. Accordingly, the relic (anti-)neutrinos can expand freely, preserving the form of $f_{\nubar}(P,\Tnu)$ in Eq.~(\ref{eq:FDdistri}). \footnote{Strictly speaking, the neutrino distribution is not an equilibrium one, since more precisely the Fermi factor in Eq.~(\ref{eq:FDdistri}) reads $\exp(\sqrt{(P/\Tnu)^2+(\mnu(T_{\nu_d})/T_{\nu_d})^2})$~\cite{Peccei:2004sz}. However, since $T_{\nu_d}\sim 1$ MeV and accordingly $\frac{\mnu(T_d)}{T_d}<10^{-6}$ for $\mnu(T_d)<1$ eV, the bulk of the neutrino distribution with $\frac{P}{\Tnu}>\frac{\mnu(T_d)}{T_d}$ is not affected by the strongly suppressed mass correction (neither in the case of constant mass neutrinos nor for MaVaNs) and can thus well be characterized by an equilibrium temperature $\Tnu$ according to Eq.~(\ref{eq:FDdistri}).} 

In order to express the cross-section $\sigma_{\nu\nubar} (P,E)$ of the $\nu\nubar$ annihilation process it is convenient to introduce the parameter $\xi=\Gamma^2_Z/M^2_Z\ll 1$. It describes the square of the quotient of the total width for $Z$ decaying to fermion pairs, $\Gamma_Z=2.4952$ GeV~\cite{Eidelman:2004wy}, and the mass of the $Z$, $M_Z=91.1876$ GeV~\cite{Eidelman:2004wy}. Accordingly, $\sigma_{\nu\nubar} (P,E)$ can be written in the following form~\cite{D'Olivo:2005uh},
\begin{equation}
\sigma_{\nu\nubar}(P,E)=
\frac{G_F}{\sqrt{2}}\frac{\Gamma_Z M_Z}{2 E^2} \frac{1}{P
E_P}\int_{s_-}^{s_+} ds \
\frac{s(s-2\mnu^2)}{\left(s-M^2_Z\right)^2+\xi s^2},\label{eq:sigma} 
\end{equation}
where $G_F=1.166\,37\times 10^{-5}\, {\rm GeV}^{-2}$ is the Fermi coupling constant and $E_p=\sqrt{P^2+\mnu^2}$ is the energy of the relic neutrino. Furthermore, $s$ is the square of the center-of-mass energy of the neutrino-anti-neutrino system. Using $\sqrt{E^2+\mnu^2}\simeq E$ for an ultra-relativistic \UHEnu one arrives at the following expression for $s$,
\be
s = 2\mnu^2 +2E(E_p-P\cos\theta),\label{eq:s}
\ee
where $\theta$ characterizes the direction of the relic anti-neutrino with respect to the line of flight of the incident \UHEnu in the center-of-mass system. Accordingly, for fixed $P$ and $E$ the integration over $s$ corresponds to the angular integration. As a consequence, the limits of the integral in Eq.~(\ref{eq:sigma}) take the values,
\be
s_\pm = 2\mnu^2 + 2E(E_p\pm P),\label{eq:spm}
\ee  
corresponding to $\cos\theta=\mp 1$.

Note that, following Ref.~\cite{D'Olivo:2005uh}, in Eq.~(\ref{eq:sigma}) the energy dependence of the $Z$ boson width $\Gamma(s)$ in the $Z$ propagator due to higher order corrections~\cite{Bardin:1988xt,Bardin:1989qr} has been taken into account. In the region of the resonance where the $s$ dependence is crucial, $\Gamma(s)$ can well be approximated by the linear relation~\cite{Bardin:1989qr,Bardin:1988xt}
\be
\label{linear}
\Gamma(s)=\frac{\Gamma_Z}{M^2_Z} s=\sqrt{\xi}\frac{s}{M_Z},\,\,\mbox{where}\,\,\Gamma_Z={\rm {const.}}.
\ee
The integral in Eq.~(\ref{eq:sigma}) for the cross-section can be solved analytically. Since the MaVaN mass $\mnu(z)$ is a decreasing function of redshift, it has its maximal value today which corresponds to the mass of a constant mass neutrino. Accordingly, for both constant mass neutrinos and MaVaNs one can exploit that $\mnu \ll M_z,E$ for all redshifts such that one finally gets the following expression for the cross-section~\cite{D'Olivo:2005uh}\footnote{The neglection of the neutrino mass in Eq.~(\ref{eq:sigmanubar}) leads to a spurious singularity in the integrand of Eq.~(\ref{eq:gammasigma}) in a region of the relic neutrino momenta where the integral is supposed to be negligible. This can be cured by an appropriate limitation of the integration interval.}

\bea
\hspace{-2cm}\sigma_{\nu\nubar}(P,E)&=& \frac{2\sqrt{2}G_F\Gamma_Z M_Z}{2EE_p}\left\{ \frac{1}{1+\xi} \right. \nonumber\\
&&\nonumber\\
&+& \frac{M_Z^2}{4EP(1+\xi)^2}\ln\left(\frac{(1+\xi)4E^2(E_p + P)^2 -4M^2_ZE(E_p+P)+M_Z^4}{(1+\xi)4E^2(E_p - P)^2 -4M^2_ZE(E_p-P)+M_Z^4}\right)\nonumber\\
&& \nonumber\\
&+&\frac{1-\xi}{(1+\xi)^2}\frac{M_Z^3}{4EP\Gamma_Z}\left[\arctan\left(\frac{2E(1+\xi)(E_p+P)-M_Z^2}{\Gamma_Z M_Z}\right) \right.\nonumber\\
&& \nonumber\\
&&\left.\left.-\arctan\left(\frac{2E(1+\xi)(E_p-P)-M_Z^2}{\Gamma_Z M_Z}\right)\right]\right\}.\nonumber\\\label{eq:sigmanubar}
\eea

The calculation of the damping defined by Eq.~(\ref{eq:gammasigma}) in combination with Eq.~(\ref{eq:sigmanubar}) includes the full thermal background effects. To allow for a comparison of our findings to published results, in the following, we will summarize two common approximations used in the literature. They result from averaging over the neutrino momenta~\cite{Barenboim:2004di} or from completely neglecting the relic neutrino motion~\cite{Weiler:1982qy,Weiler:1983xx,Roulet:1992pz,Eberle:2004ua}. As will become apparent in the next section, in the neutrino mass range favored by the supersymmetric MaVaN model (cf. Sec.~\ref{sec:MaVaNs}) their applicability is quite limited. However, the assumption of relic neutrinos at rest will later on remove the thermal distortion of the absorption dips and thus allow a deeper insight into the characteristic features caused by the mass variation $\mnui(z)$ described in Sec.~\ref{sec:MaVaNs}.

The weaker approximation for $\sigma_{\nu\nubar}(P,E)$ commonly used in the literature is obtained by approximating the mean value theorem. The factor $(s-2\mnu^2)$ in the integrand of Eq.~(\ref{eq:sigma}) takes the role of the weight function and is integrated over, whereas the residual part of the integrand is -- as an approximation -- evaluated at the midpoint of the integration interval $\bar{s} \simeq 2EE_p=2 E\sqrt{P^2+\mnu^2}$. Accordingly, the cross-section takes the following form
\begin{equation}
\label{eq:barsigmabar}
\bar{\sigma}_{\nu\nubar} (E,P)= \bar{\sigma}_{\nu\nubar} (\bar{s})=2\sqrt{2} G_F \Gamma_Z M_Z \ \   
\frac{\bar{s}}{\left(\bar{s} - M_Z^2\right)^2 + \xi \bar{s}^2},
\end{equation}
with $\xi=\Gamma^2_Z/M^2_Z$. Let us point out again that due to higher order corrections a precise handling of the $Z$ propagator near the resonance~\cite{Bardin:1989qr,Bardin:1988xt} leads to an energy dependence of the $Z$ boson width $\Gamma(\bar{s})$ (cf. Eq.~(\ref{linear})).
Note that in earlier work on neutrino-absorption (e.g. \cite{Weiler:1982qy,Weiler:1983xx,Roulet:1992pz,Eberle:2004ua}) this correction has not been made but the lowest order (simple Breit-Wigner) form for the cross-section $\bar{\sigma}_{\nu\nubar} (E,P)$ has been used. We would like to stress that by the averaging procedure,
which leads to the approximation for $\sigma_{\nu\nubar}$ in Eq.~(\ref{eq:barsigmabar}), part of the angular information gets lost. This results in an underestimation of the thermal spread of the absorption dips~\cite{D'Olivo:2005uh}.

Let us now come to the second, more radical approximation: In earlier work on cosmic neutrino absorption (e.g. ~\cite{Roulet:1992pz,Eberle:2004ua}) it was assumed that the relic neutrinos are at rest~\cite{Roulet:1992pz,Eberle:2004ua}, thereby switching off all thermal background effects. The corresponding cross-section can be recovered from the full expression in Eq.~(\ref{eq:sigma}) by taking the limit $P\rightarrow 0$ or from eq.~(\ref{eq:barsigmabar}) by setting $P=0$ such that $\bar{s}=s_0=2E\mnu$. In this case, the remaining integral over $P$ in Eq.~(\ref{eq:gammasigma}) reduces to the neutrino number density $n_{\nu}(z)=n_{\nubar}(z)=n_{\nu 0}(1+z)^3$. Accordingly, the damping $\gamma_{\nu\nubar}$ takes the following form
\be
\label{eq:gammaRoulet}
\hspace{-2cm}\gamma^0_{\nu\nubar}(E) = \bar{\sigma}_{\nu\nubar}(s_0)\ n_\nu(z)=2\sqrt{2} G_F \Gamma_Z M_Z\ \frac{2 E m_{\nu}}{4(1+\xi)E^2 m^2_{\nu} - 4M_Z^2E m_{\nu} +M_Z^4} \ n_{\nu}. 
\ee
As will become apparent in the next section, by assuming the neutrinos to be at rest one neglects two conspiring effects on the damping which become more important with decreasing ratio $\mnu/\Tnu$~\cite{D'Olivo:2005uh}: 

On the one hand, the full cross-section $\sigma_{\nu\nubar}(P,E)$ in Eq.~(\ref{eq:sigmanubar}), which governs the damping, depends on $E_p=\sqrt{P^2+\mnu^2}$. As a consequence, the peak of the cross-section for a thermal bath of relic anti-neutrinos at $E_{i}^{\res}$~\cite{Barenboim:2004di},
\be
\label{angularEres}
E_{i}^{\res}=\frac{M_Z^2}{2(\sqrt{P_i^2+\mnui^2}-P_i\cos\theta)},
\ee
actually lies at lower energies than the one of $\bar{\sigma}_{\nu\nubar}(s_0)$ for relic anti-neutrinos at rest: the energy $E_{0,i}$ reduces to
\be
\label{bareEres}
E_{i}^{\res}=\frac{M^2_Z}{2\mnui}.
\ee
Note that in the case of MaVaNs the masses $\mnui$ are functions of redshift, $\mnu(z)$, and therefore introduce a $z$ dependence into the resonance energies. Thus, they only coincide with the respective constant mass ones for $z=0$ and $\mnu(0)=\m0$. We will discuss the consequences for the absorption features in detail in the next section.

As indicated by Eq.~(\ref{angularEres}), the effect of the relic neutrino momenta $P$ becomes significant for small neutrino masses, according to Ref.~\cite{D'Olivo:2005uh} for $\mnu \le 0.01$ eV. Furthermore, $\bar{\sigma}_{\nu\nubar}(s_0)$ overestimates the peak height of the full expression and cannot account for the broadening of $\sigma_{\nu\nubar}(P,E)$ for increasing relic neutrino momentum $P$. 

On the other hand, the thermal distribution of the relic neutrinos which gives rise to a Fermi momentum smearing of the cross-section is totally neglected. In the full expression the damping $\gamma_{\nu\nubar}$  results from the integration over all neutrino momenta, where the weight factor $P^2 f_{\nubar}(P,\Tnu)$ selects relic neutrino momenta $P$ of the order of $\Tnu$. Accordingly, Eq.~(\ref{eq:gammaRoulet}) overestimates the damping efficiency with respect to the full expression defined by Eq.~(\ref{eq:sigmanubar}) and Eq.~(\ref{eq:gammasigma}). As we will see in the next section the realistic description of neutrino-absorption leads to less well defined absorption features spread over a larger range of \UHEnu energies than in the idealized scenario which neglects any thermal effects. These discrepancies increase with decreasing neutrino mass. 

\section{Survival Probability of Extremely High-Energy Cosmic Neutrinos Traversing the Cosmic Neutrino Background \label{sec:SurvivalProb}} %

The relevant quantity to be discussed in this section is the survival probability $P_{\nu_{\alpha}}$ of extremely high-energy cosmic neutrinos $\nu_{\alpha}$ of flavor $\alpha=e,\mu,\tau$ traveling through the \CnuB to us. It is governed by the damping rate $\gamma_{\nu\nubar}$ introduced in the last section and it determines, folded with the respective \UHEnu source emissivity distribution ${\cal{L}}_{\nu_i}$, the resulting neutrino spectra to be observed on earth, which are treated in Sec.~\ref{sec:AbsorptionDips}. 

The main goal of this section is to work out the characteristic differences in the shape of the absorption dips in the \UHEnu survival probabilities which arise from considering the neutrino masses to be dynamical quantities $\mnui(z)$ instead of constant parameters. After presenting our results and pointing out the generic differences, we will have a closer look at the MaVaN absorption features. As motivated in the last section, for the purpose of gaining more physical insight, we will disentangle the different influences which define their shape. First of all, we will study the impact of the mass variation $\mnui(z)$ as well as of the cosmic redshift caused by the expansion of the universe. To this end, we will initially switch off any thermal background effects by assuming the relic neutrinos to be at rest and compare the results for MaVaNs to those of constant mass neutrinos. 

Nonetheless, we would like to stress again that due to the low neutrino mass scale required in the MaVaN model under consideration (cf. Sec.~\ref{sec:MaVaNs}) only the full treatment of the background effects can serve as a test for Neutrino Dark Energy.

In our calculation we make the standard simplifying assumption that the \UHEnu source switched on at a distinct redshift $z_s$ in the past. As concerns plausible \UHEnu sources, in the following we would like to mention the most relevant classes as well as the corresponding typical \UHEnu emission redshifts $z_s$. 

As a first possibility, \UHEnu are assumed to originate from pion decays, where the latter either are produced in inelastic $pp$ or $p\gamma$ interactions. Those astrophysical acceleration sites (bottom-up mechanism), notably active galactic nuclei (AGN) and gamma-ray bursters, have source positions $z_s$ of a few (e.g.~\cite{Protheroe:2004rt,Torres:2004hk}). The conjectured energies of cosmic neutrinos produced by these astrophysical acceleration sites in the case of shock acceleration are $E_{\max}\lwig 10^{11}-10^{12}$ GeV~\cite{Protheroe:2004rt,Torres:2004hk}. However, even higher energies are possible in proposed non-shock acceleration mechanisms, such as unipolar induction, acceleration in strong magnetic waves in plasmas (wakefields)~\cite{Chen:2002nd}, or by magnetic recombination in the vicinity of massive black holes~\cite{Li:1998yg,Kronberg:2001st} (see~\cite{Protheroe:2004rt} for a recent review).

As a second possibility, extremely energetic cosmic neutrinos with energies above $10^{12}$ GeV may be generated in the decomposition of so-called topological defects (top-down scenarios) into their constituent particles. Topological defects are predicted to originate from symmetry breaking phase transitions immediately after (hybrid) inflation (see~\cite{Sakellariadou:2005wy} for a recent review). In particular, cosmic string formation is highly generic in Supersymmetric Grand Unified Theories (SUSY GUTs) (see~\cite{Sakellariadou:2006qs} and references therein). Those topological defects produce super-heavy quanta generically denoted as $X$ particles (often heavy Higgs or gauge bosons) with masses $m_X\sim 10^{12}-10^{16}$ GeV. Those $X$ particles rapidly decay into stable Standard Model (SM) particles, also releasing extremely energetic neutrinos~\cite{Bhattacharjee:1991zm,Bhattacharjee:1998qc} with energies up to $\sim 0.5\, m_X$~\cite{Ringwald:2005wa}. For those exotic, non-accelerator sources, $z_s$ can be as high as the epoch of light neutrino decoupling, $z_s\sim \rm{\cal{O}}(10^{10})$~\cite{Gondolo:1991rn}. 

In our analysis we take resonant $Z$-production caused by the interaction with the \CnuB as the only source of attenuation of the propagating \UHEnu. This approximation is well justified in the energy regions of the absorption dips which we are focusing on in our investigation~\cite{Eberle:2004ua}. Accordingly, the survival probability of an extremely high-energy cosmic neutrino $\nu_{i}$ with $i=1,2,3$ injected at redshift $z_s$ is given by,
\be
\label{SurvivalProb}
P_{\nu_i}(E_0,z_s)=\exp\left[-\int^{z_s}_{0}\frac{dz}{H(z)(1+z)}\gamma_{\nu\nubar}\big(E_0(1+z)\big)\right],\,\,i=1,2,3,
\ee
where the integral in the exponential, which governs the survival probability, is called the optical depth (or the opacity). It contains the product of the propagation distance $dr=dz/[(1+z)H(z)]$ and the damping rate $\gamma_{\nu\nubar}(E)$ defined in Eq.~(\ref{eq:gammasigma}) with $E=E_0\,(1+z)$, which is integrated over all redshifts from the present time up to the emission redshift $z_s$. Moreover, in a universe with negligible radiation component the evolving Hubble factor is given by 
\be
H(z)=H_0\big(\Omega_M(1+z)^3+\Omega_k(1+z)^2+\Omega_{\Lambda}\big)^{1/2}. 
\ee
As suggested by recent observations~\cite{Eidelman:2004wy}, we take a present day matter density $\Omega_M=0.3$, a curvature density $\Omega_k=0$ and vacuum energy density $\Omega_{\Lambda}=0.7$ as corresponds to a Lambda Cold Dark Matter ($\Lambda$CDM) universe. Note that this specific form for $H(z)$ also applies to the MaVaN scenario under consideration: according to Sec.~\ref{sec:MaVaNs}, the neutrino dark energy density $\Omega_X$ redshifts with an equation of state $\omega \sim -1$~\cite{Fardon:2005wc} and therefore behaves very much like a cosmological constant $\Lambda$, $\Omega_X \sim \Omega_{\Lambda} \sim {\rm{const.}}$. 

We will express our results for the survival probabilities in terms of the propagating neutrino flavors $\nu_{\alpha}$ according to,
\bea
\label{SurvivalProbflavor}
P_{\nu_{\alpha}}&=&\sum\limits_i^{}|U_{\alpha i}|^2 P_{\nu_i},\,\,\mbox{with}\\
i&=&1,2,3\,\,\mbox{and}\,\,\alpha=e,\mu,\tau,\nonumber
\eea
where the absolute square of the leptonic mixing matrix elements $U_{\alpha i}$ relates the neutrino flavor components $\nu_{\alpha}$ to the mass eigenstates $\nu_i$. Note that since the mixing matrix element $|U_{e 3}|\ll 1$, the absorption dip produced by the heaviest mass eigenstate will not be visible in the case of $P_{\nu_e}$. However, apart from this exception, the flavor survival probabilities to be discussed in the following exhibit absorption dips at the respective resonance energies of the mass eigenstates $\nu_i$, for $i=1,2,3$.

\begin{figure}
\vspace* {0.0in}
\begin{center}
\includegraphics*[bbllx=20pt,bblly=119pt,bburx=550pt,bbury=744pt,height=9cm,width=7.7cm]{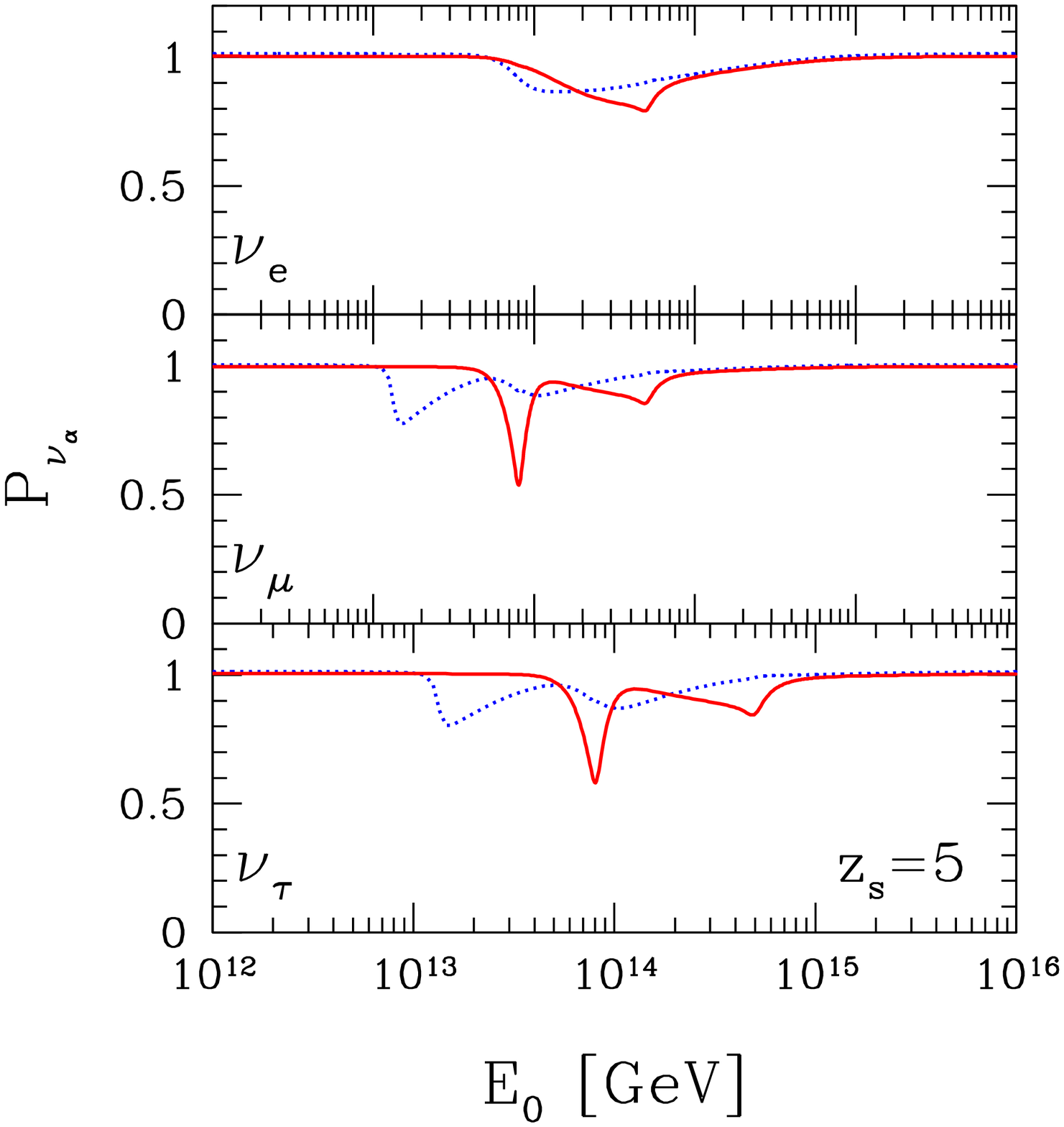}
\includegraphics*[bbllx=20pt,bblly=119pt,bburx=550pt,bbury=744pt,height=9cm,width=7.7cm]{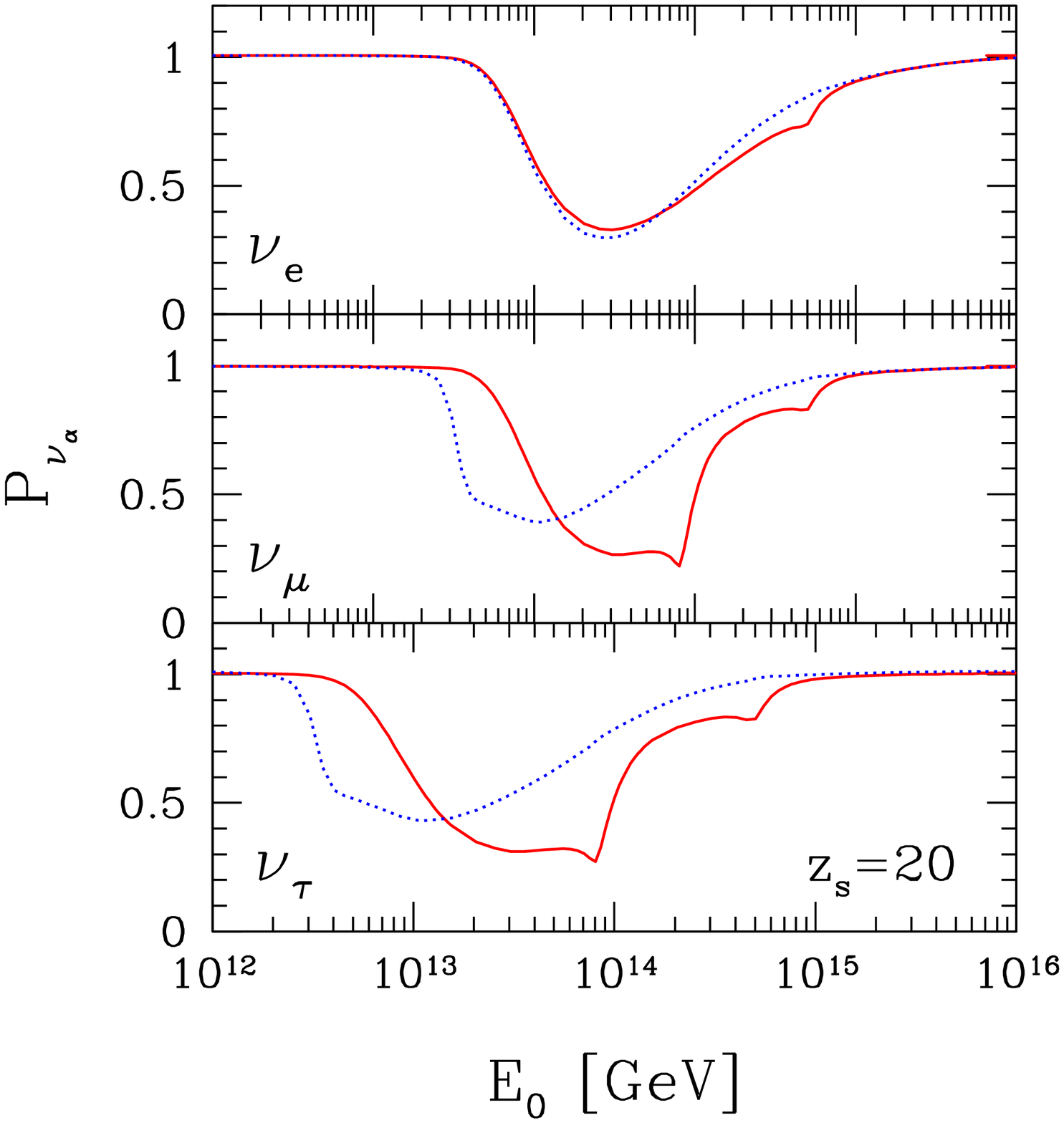}
\caption[]{Flavor survival probability $P_{\nu_{\alpha}}$, $\alpha=e,\mu,\tau$ with all thermal background effects included and integrated back to $z_s=5$ (left panel) and $z_s=20$ (right panel), respectively, for a normal neutrino mass hierarchy with $m_{\nu_{0,1}}=10^{-5}$ eV and varying neutrino masses (solid lines) as well as constant neutrino masses (dotted lines).}
\label{fig3}
\end{center}
\end{figure}  

\begin{figure}
\vspace* {0.0in}
\begin{center}
\includegraphics*[bbllx=20pt,bblly=119pt,bburx=590pt,bbury=746pt,height=9cm,width=8.3cm]{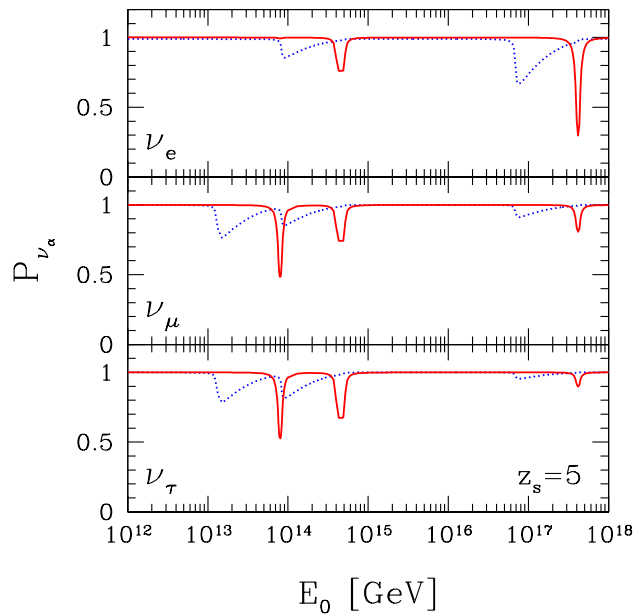}
\caption[]{Approximated flavor survival probability for $P_{\nu_{\alpha}}$, $\alpha=e,\mu,\tau$, which assumes the target relic-neutrinos to be at rest, $P=0$, after an integration back to $z_s=5$, for a normal neutrino mass hierarchy with $m_{\nu_{0,1}}=10^{-5}$ eV and varying neutrino masses (solid lines) as well as constant neutrino masses (dotted lines) plotted as a function of their energy $E_0$ at earth.}
\label{fig4}
\end{center}
\end{figure}

\begin{figure}
\vspace* {0.0in}
\begin{center}
\includegraphics*[bbllx=20pt,bblly=119pt,bburx=550pt,bbury=744pt,height=9cm,width=7.7cm]{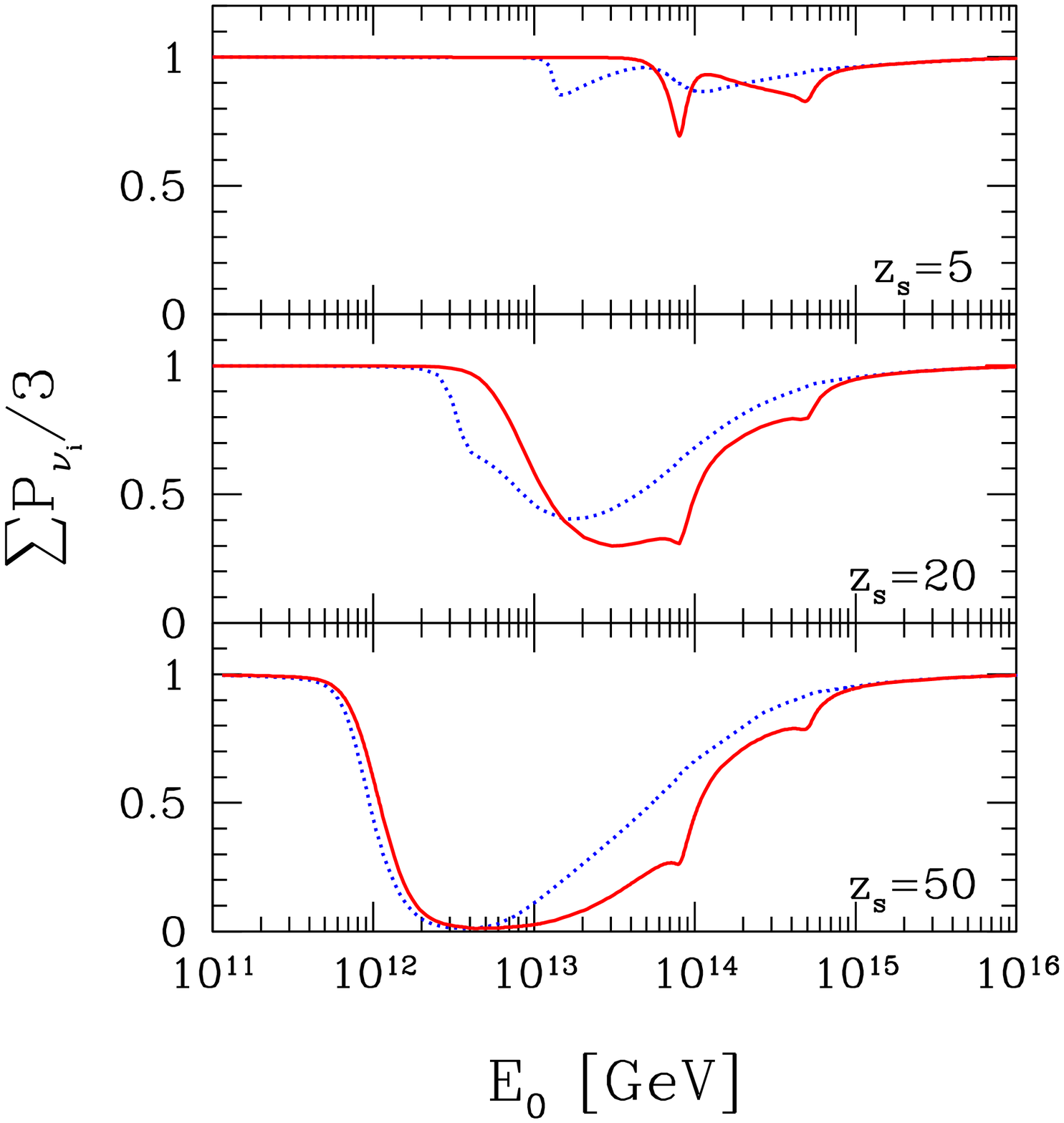}
\includegraphics*[bbllx=20pt,bblly=119pt,bburx=550pt,bbury=744pt,height=9cm,width=7.7cm]{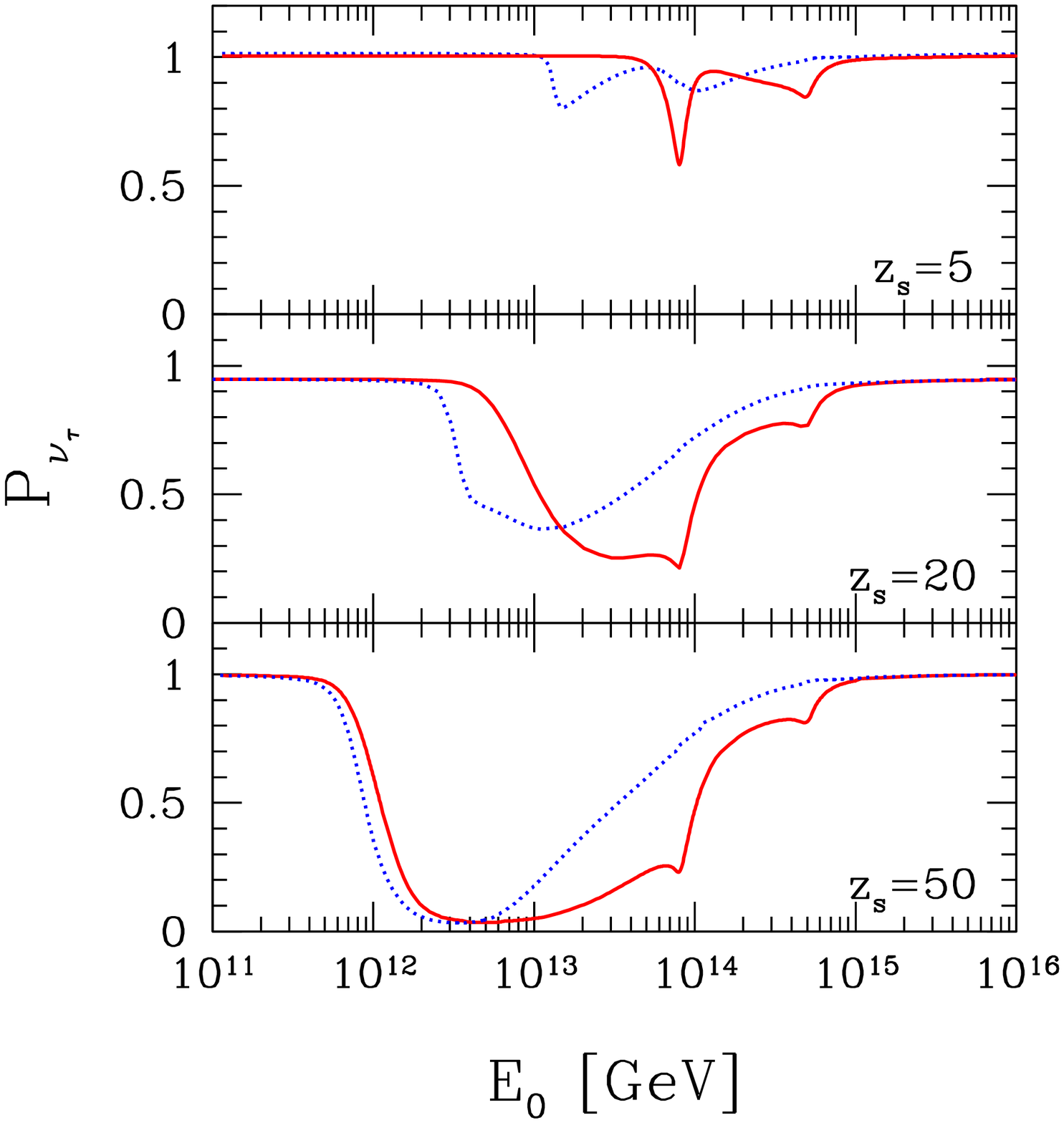}
\caption[]{The normalized sum of the survival probabilities, $\frac{1}{3}\sum P_{\nu_i}$, and the flavor survival probability $P_{\nu_{\tau}}$, respectively, including all thermal background effects, for a normal neutrino mass hierarchy with $m_{\nu_{0,1}}=10^{-5}$ eV and varying neutrino masses (solid lines) as well as constant neutrino masses (dotted lines) plotted as a function of their energy $E_0$ at earth for $z_s=5$, $z_s=20$ and $z_s=50$.}
\label{fig5}
\end{center}
\end{figure}

Let us begin our analysis by comparing our results for the flavor survival probabilities defined in Eq.~(\ref{SurvivalProb}) and Eq.~(\ref{SurvivalProbflavor}), each with varying and constant mass, including all thermal background effects according to Eq.~(\ref{eq:gammasigma}) and Eq.~(\ref{eq:sigmanubar}). Throughout this section we assume a normal neutrino mass hierarchy, where the masses to take values according to Eq.~(\ref{m1}) -- Eq.~(\ref{m3}) and the mass variation is given in Fig.~\ref{fig2} (cf. Sec.~\ref{sec:MaVaNs}). For numerical values of $|U_{\alpha i}|$ we take~\cite{Altarelli:2004cp,Li:2004dn}~\footnote{Apparently, unitarity is not exactly fulfilled for the best fit values in Eq.~(\ref{mixing}). This leads to a small shift in the flavor survival probabilities $P_{\nu_{\alpha}}$, which we have compensated for in Fig.\ref{fig3} -- Fig.\ref{fig5} in such a way that $P_{\nu_{\alpha}}=1$ outside the energy regions of the absorption dips produced by the mass eigenstates $\mnui$.}
\be
\label{mixing}
|U_{\alpha i}|=\left(\begin{array}{ccc}0.84\,\, 0.54\,\,\, 0.08\\0.44\,\,\, 0.56\,\,\, 0.72\\0.32\,\,\, 0.63\,\,\, 0.69\end{array}\right).
\ee

Let us start by considering the flavor survival probabilities $P_{\nu_{\alpha}}$ for an emission redshift of $z_s=5$ which are plotted in Fig.~\ref{fig3} on the left hand side. Apparently, both in the case of varying and constant neutrino masses, the absorption features produced by the lighter two neutrino mass eigenstates are merged together. However, for MaVaNs, the absorption dips produced by the heavier two neutrino mass eigenstates exhibit sharp spikes at the respective resonance energies $E^{res}_{0,i}=\frac{M^2_Z}{2m_{\nu_{0,i}}}$ with $i=2,3$ instead of being washed out and distorted to lower energies as in the constant mass case. In principle, by this means, the neutrino masses $m_{\nu_{0,i}}$ for $i=2,3$ could be directly inferred from the respective spike positions of the MaVaNs dips in the low redshift regime. 

Altogether, the MaVaN absorption dips are much deeper and narrower in comparison to the constant mass features. In addition, the respective minimum positions are shifted to higher energies by almost an order of magnitude with respect to the corresponding constant mass dips.  

As demonstrated by Fig.~\ref{fig3} on the right hand side, for an increased emission redshift $z_s=20$, the absorption features are considerably deeper than for $z_s=5$. In addition, for each neutrino flavor the dips have merged, both in the case of varying and constant neutrino mass. As we learned in the last section, this effect of the thermal motion has increased with $z$, since the thermal bath of relic neutrinos was hotter at earlier times.  

For MaVaNs, the characteristic narrow spikes at the resonance energies $E_{0,i}^{\res}$ with $i=2,3$ are less pronounced than for $z_s=5$ and also suffer a distortion towards lower energies. Nevertheless, for $\nu_{\mu}$ and $\nu_{\tau}$, they remain well distinguishable from the respective constant mass dips. As in the case of $z_s=5$ the absorption features are clearly shifted to higher energies and exhibit substantially deeper dips. 

The characteristic absorption features produced by the mass variation can be worked out by separating the different influences on the MaVaN absorption dips. Let us for this purpose assume the relic neutrinos to be at rest, in order to eliminate any thermal background effects on the MaVaN survival probabilities. By doing so, we are left with the combined effects of the cosmic redshift and the mass variation. Let us first of all consider the former effect which is present both for MaVaNs as well as for constant mass neutrinos. It originates in the expansion of the universe and manifests itself in an energy loss of \UHEnu's of energy $E$ according to $E_0=E/(1+z)$, where $E_0$ is the \UHEnu energy to be measured at earth. Accordingly, the survival probability $P_{\nu_i}$ of an \UHEnu is reduced, as long as somewhere on its way to us ($z_s\ge z\ge 0$) it has the right amount of energy,
\be
\frac{E_{0,i}^{\res}}{(1+z_s)}\le E_0 \le E_{0,i}^{\res},
\label{Eres}
\ee
to annihilate resonantly with a relic anti-neutrino (whereas for all other energies the \CnuB is transparent for the \UHEnu such that $P_{\nu_{\alpha}}=1$). As a consequence, the effect of cosmic redshift is observable in a broadening of the \UHEnu absorption dips.

Let us stress that in addition to this cosmological effect in the case of MaVaNs, the variation of the neutrino masses $\mnui(z)$ causes a redshift dependence of the respective resonance energies $E_i^{\res}(z)$ as already mentioned in the last section. To be more concrete, the masses $\mnui(z)$ at redshift $z$ determine the corresponding resonance energies to be $E_i^{\res}(z)=M^2_Z/2 \mnui(z)$ in the rest system of the relic neutrinos which only coincide with $E_{0,i}^{\res}=M^2_Z/2 m_{\nu_{0,i}}$ for $z = 0$.

In Fig.~\ref{fig4} we plot the respective survival probabilities of the neutrino flavors with varying and constant masses, neglecting the relic neutrino momenta and integrating back to $z_s=5$. As expected from the discussion in the last section, the thermal spread of the absorption features provoked by the relic neutrino motion is removed such that the dips do not merge. Instead, for constant mass neutrinos, the absorption features in Fig.~\ref{fig4} are only subject to the broadening caused by the cosmic redshift and span the energy interval specified in Eq.~(\ref{Eres}). 

In striking contrast, the absorption dips produced by the MaVaN mass eigenstates exhibit sharp minima at the resonance energies $E_{0,i}^{\res}=\frac{M^2_Z}{2\m0}$ with $i=1,2,3$ showing no spread towards lower energies. Thus, interestingly, these MaVaN dips look like those of constant mass neutrinos in a non-expanding universe. Actually, it turns out that the mass induced redshift dependence of the resonance energies $E_i^{\res}(z)=\frac{M^2_Z}{2\mnui(z)}$, $i=1,2,3$, compensates for the energy loss of the \UHEnu due to cosmic redshift in the limit of low redshifts. This can be understood by remembering that the approximation in Eq.~(\ref{LR}) gives a good estimate for the redshift dependence of the neutrino masses $m_{\nu_i}(z)$ in the low redshift regime (cf. Fig.~\ref{fig2}). In this limit one arrives at the following functional dependence of the resonance energies on $z$, 
\be
E_i^{\res}(z)=\frac{M^2_Z}{2 \mnui(z)}=E_{0,i}^{\res}(1+z).
\ee
In turn, the resonance energy $E_i^{\res}(z)=E_{0,i}^{\res}(1+z)$ of an \UHEnu at redshift $z$ corresponds to the redshifted energy measured at earth,
\be
\label{EresPeak}
\frac{E_i^{\res}(z)}{(1+z)}=E_{0,i}^{\res},\,\,\mbox{with}\,\,i=2,3,
\ee
Accordingly, the annihilation of an \UHEnu on the \CnuB at any given redshift $z_s\ge z \ge 0$ always leads to an absorption peak at $E_{0,i}^{\res}$. In other words, in this approximation the square of the center-of-mass energy, $s_{0,i}=2\mnui(z)E_0(1+z)=2m_{\nu_{0,i}}E_0$, $i=1,2,3$, becomes redshift independent. Correspondingly, the undistorted shape of the annihilation cross-section $\bar{\sigma}_{\nu\nubar}(s_{0,i})$ (weighted with the neutrino density per unit redshift and integrated over $z$) gets projected on the sky. 

Note that even for higher redshifts the appropriate mass-redshift approximation from Eq.~(\ref{HR}) leads to the following redshift dependence of $E_i^{\res}(z)$,
\be
E_i^{\res}(z)=\frac{M^2_Z}{2 m_{\nu_i}(z)}\propto (1+z)^{1/2}\,\,\mbox{with}\,\,i=1,2,3.
\ee
Apparently, the neutrino mass variation still partially counterbalances the effect of the cosmic redshift at high redshifts.

In summary, all of the respective MaVaN absorption lines can be distinguished from those of constant mass neutrinos, since the redshift distortion is much less pronounced in any case. We would like to point out that this is a generic feature of any standard MaVaN scenario where the neutrino mass is a decreasing function of redshift.

We are now in a position to complete the interpretation of the MaVaN features in Fig.~\ref{fig3} which result from combining all effects on the absorption features including the thermal ones. As already mentioned in the last section, the Fermi-weight factor $P^2 f_{\nubar}(P,T)$ entering the damping integral in Eq.~(\ref{eq:gammasigma}) selects momenta $P$ of the order of the \CnuB temperature $\Tnu(z)$. Accordingly, the ratio $\mnui/\Tnu(z)$ is a measure for the relevance of the mass and its probable evolution with respect to the temperature effects. Let us in the following discuss the absorption features which are completely determined by the thermal effects. In these cases the absorption lines were produced by relativistic neutrinos, since $\mnui(z)/\Tnu(z)\ll 1$. Firstly, for $\nu_{e}$ the absorption lines of MaVaNs and constant mass neutrinos are similar (cf. Fig.~\ref{fig3}). This can be understood by recalling that $\nu_e$ is mostly composed of the lightest mass eigenstate for which mass effects neither today nor in the past have played any role, since ($m_{\nu_{0,1}}/T_{\nu_0}\ll 1$). Secondly, as opposed to the case of $z_s=5$, the MaVaN absorption features for $z_s=20$ are distorted to much lower energies (cf. Fig.~\ref{fig3}). In addition, for MaVaNs, the low energy ends of the dips for $\nu_{\mu}$ and $\nu_{\tau}$ have the same shape as the one of $\nu_e$. The reason is, that the absorption lines in this energy region stem from absorptions at high $z<z_s$ (as indicated by Eq.~(\ref{Eres}) in absence of any thermal effects), where all neutrino masses still were negligible with respect to the temperature. As a consequence, they are clearly distinguishable from the corresponding ones of constant mass neutrinos, which apparently are already non-relativistic in the same energy region. This is due to the fact that while the temperature rises with increasing $z$, only the MaVaN masses $\mnui(z)$ evolve and become lighter. Thus, MaVaNs generically turn non-relativistic much later than constant mass neutrinos. At energies above this transition from the non-relativistic to the relativistic regime, the variation of the heavier two neutrino masses is not washed out by the temperature effects. Therefore, it leads to sharp and thus deep absorption minima at the respective resonance energies according to Eq.~(\ref{EresPeak}) (cf. Fig.~\ref{fig3}), without and with transition of the regimes, respectively).

As a conclusion we have learned that the characteristic effects of the neutrino mass variations in the case of the heavier two MaVaNs become apparent in the higher energy regions of the absorption dips, where the MaVaNs are still non-relativistic. However, also the low energy end of the absorption dips differs as long as the MaVaNs are relativistic and the constant mass neutrinos have already turned non-relativistic.

The next section will deal with realistic neutrino fluxes to be measured by neutrino observatories. In general, a flavor tagging at extremely high energies seems unlikely apart from exceptions (e.g. for particular flavors), whereas all observatories will be sensitive to the flavor summed \UHEnu fluxes $\sum F_{\nu_{\alpha}}$ with $\alpha=e,\mu,\tau$. Accordingly, we will base our final discussion on the totalized fluxes $\sum F_{\nu_{\alpha}}$. In addition, we will include our results for the $\nu_{\tau}$ whose identification will at best be feasible by the LOFAR radio telescope~\cite{Private}. For this purpose, for emission redshifts $z_s=5$, $z_s=20$ and $z_s=50$ in Fig.~\ref{fig5} we collect both the resulting normalized sum of the survival probabilities, $\frac{1}{3}\sum P_{\nu_i}$ which governs the $\sum F_{\nu_{\alpha}}$ as well as the respective $P_{\nu_{\tau}}$. Note that emission redshifts of $z_s=50$ (and much higher) will contribute to the \UHEnu fluxes which result from the decomposition of super-heavy particles produced by topological defects (top-down sources). Apparently, also for emission redshifts of this order, the characteristic differences between the respective absorption features of varying and constant mass neutrinos, which we have discussed above, are still visible. Furthermore, the absorption dips get considerably deeper with increasing emission redshift $z_s$.

\section{Absorption Dips in Realistic \UHEnu Spectra -- Test of Neutrino Dark Energy?  \label{sec:AbsorptionDips}}                                                                           %

So far, in our analysis we have concentrated on the determination and the comparison of the survival probabilities $P_{\nu_{\alpha}}$ of mass varying and constant mass \UHEnu's traversing the \CnuB, where the $P_{\nu_{\alpha}}$ contain the physical information on the annihilation process $\nu_{\alpha} \nubar_{\alpha} \rightarrow Z$. In the following we will outline how our results can be employed to perform relic neutrino absorption spectroscopy and to what extend the latter can serve as a test for the MaVaN scenario. For this purpose, we will firstly consider astrophysical \UHEnu acceleration sites and, secondly, topological defect sources both introduced in the last section. In the latter case, we will not approximate the \UHEnu injection spectrum by a standard power-law, but go through the appropriate calculation involving fragmentation functions as discussed below.

As a starting point, we require both the detection of \UHEnu fluxes in the energy region of interest as well as the observation of absorption lines in these spectra. The \UHEnu flux $F_{\nu_{\alpha}}(E_0)$ for a neutrino of flavor $\alpha=e,\mu,\tau$ to be measured at earth is given by~\cite{Eberle:2004ua}
\be
\label{Flux}
\hspace{-2.3cm}F_{\nu_{\alpha}}(E_0)=\frac{1}{4\pi}\int\limits_{0}^{\infty}\frac{dz_s}{H(z_s)}\times\sum\limits_{\beta,j}^{}|U_{\alpha j}|^2 P_{{\nu}_j}(E_0\,(1+z_s),z_s)\,|U_{\beta j}|^2 {\cal{L}}_{\nu_{\beta}}(E_0\,(1+z_s),z_s).
\ee
The flux integral is governed by the survival probability $P_{{\nu}_j}$ defined in Eq.~(\ref{SurvivalProb}) weighted by the \UHEnu source emissivity distribution ${\cal{L}}_{\nu_{\beta}}$, which depends on the properties of the source as described in the following. On the one hand, the diffuse source emissivity ${\cal{L}}_{\nu_{\beta}}(E_0\,(1+z),z_s)$ takes into account the distribution of the sources in the universe (the activity $\eta$) and on the other hand it considers the number of neutrinos of flavor $\beta=e,\mu,\tau$ emitted by each of the sources (the injection spectrum $J_{\nu_{\beta}}$). Under the standard assumption of identical injection spectra for all sources, one can factorize the $z$ and $E$ dependence,
\be
\label{LumFac}
{\cal{L}}_{\nu_{\beta}}(z_s,E)=\eta(z_s)J_{\nu_{\beta}}(E),\,\,
\mbox{with}\,\,E=E_0(1+z_s).
\ee 

As already stated in the last section, a flavor tagging at extremely high-energies cannot be expected at all neutrino observatories. However, we can hope for the identification of the $\nu_{\tau}$ absorption lines at LOFAR~\cite{Private}, which we will therefore consider according to Eq.~(\ref {Flux}). Furthermore, in our analysis, we will consider the total flux of all neutrino flavors $\sum F_{\nu_{\alpha}}$ which can well be approximated by~\cite{Eberle:2004ua}, 
\be
\label{FluxSum}
\hspace{-1.5cm}\sum F_{\nu_{\alpha}}(E_0)\simeq \frac{1}{4\pi}\int\limits_{0}^{\infty}\frac{dz_s}{H(z_s)}\frac{1}{3}{\cal{L}}^{\rm tot}_{\nu}(E_0\,(1+z_s),z_s)\sum\limits_{j=1}^{3}P_{{\nu}_j}(E_0\,(1+z_s),z_s),
\ee
where ${{\cal{L}}^{\rm tot}_{\nu}}$ denotes the total, flavor-summed neutrino emissivity at the source and the formula holds as long as ${\cal{L}}_{\nu_{\mu}}+{\cal{L}}_{\nu_{\tau}}=2{\cal{L}}_{\nu_{e}}$. The latter is fulfilled for hadronic sources like astrophysical accelerator bottom-up sources or non-accelerator top-down sources, since in both cases the neutrinos emerge from charged pion decays such that
\be
{\cal{L}}_{\nu_{e}}:{\cal{L}}_{\nu_{\mu}}:{\cal{L}}_{\nu_{\tau}}=1:2:0.
\ee     
However, Eq.~(\ref{FluxSum}) also holds in the case of equal flavor source emissivities,
\be
{\cal{L}}_{\nu_{e}}:{\cal{L}}_{\nu_{\mu}}:{\cal{L}}_{\nu_{\tau}}=1:1:1,
\ee     
as could arise in the decays of topological defects not directly coupled to matter (e.g. mirror-matter `necklaces')~\cite{Berezinsky:1999az,Berezinsky:2002fa}.

Note that the dependence on the leptonic mixing matric elements $|U_{\alpha j}|$, present in Eq.~(\ref{Flux}), has dropped out in the expression for $\sum F_{\nu_{\alpha}}$ in Eq.~(\ref{FluxSum}) due to unitarity. 

In the next subsection we start our investigation by considering astrophysical (bottom-up) \UHEnu sources. In the subsequent subsection we continue our analysis for the case of topological defect (top-down) \UHEnu sources. 

\subsection{Astrophysical neutrino sources}                                                                           %

In the following we will discuss \UHEnu fluxes which are assumed to originate from astrophysical \UHEnu sources. In order to parameterize their source emissivity distribution ${\cal{L}}_{\nu_{\beta}}$ we employ the following standard ansatz (e.g. ~\cite{Kalashev:2002kx,Semikoz:2003wv,Eberle:2004ua}) in combination with Eq.~(\ref{LumFac}),
\bea
\label{eta}
\eta(z_s)=\eta_0(1+z_s)^n\theta(z_s-z_{min})\theta(z_{\max}-z_s),\\
\label{J}
J_{\nu_{\beta}}(E)=j_{\nu_{\beta}}E^{-\alpha}\theta(E-E_{\rm{min}})\theta(E_{\max}-E).
\eea
Throughout our analysis, we will take $z_{\rm{min}}=0$ and $E_{\rm{min}}=0$ as default values and suppose that $E_{\max}>E^{\res}_{0,i}(1+z_{\max})$ for $i=2,3$. Furthermore, we will not examine the possibility of broken power-law injection spectra, but assume the spectral index $\alpha$ to be constant in the whole energy region of interest.

\begin{figure}
\begin{center}
\includegraphics*[bbllx=20pt,bblly=221pt,bburx=570pt,bbury=673pt,height=7cm,width=7.9cm]{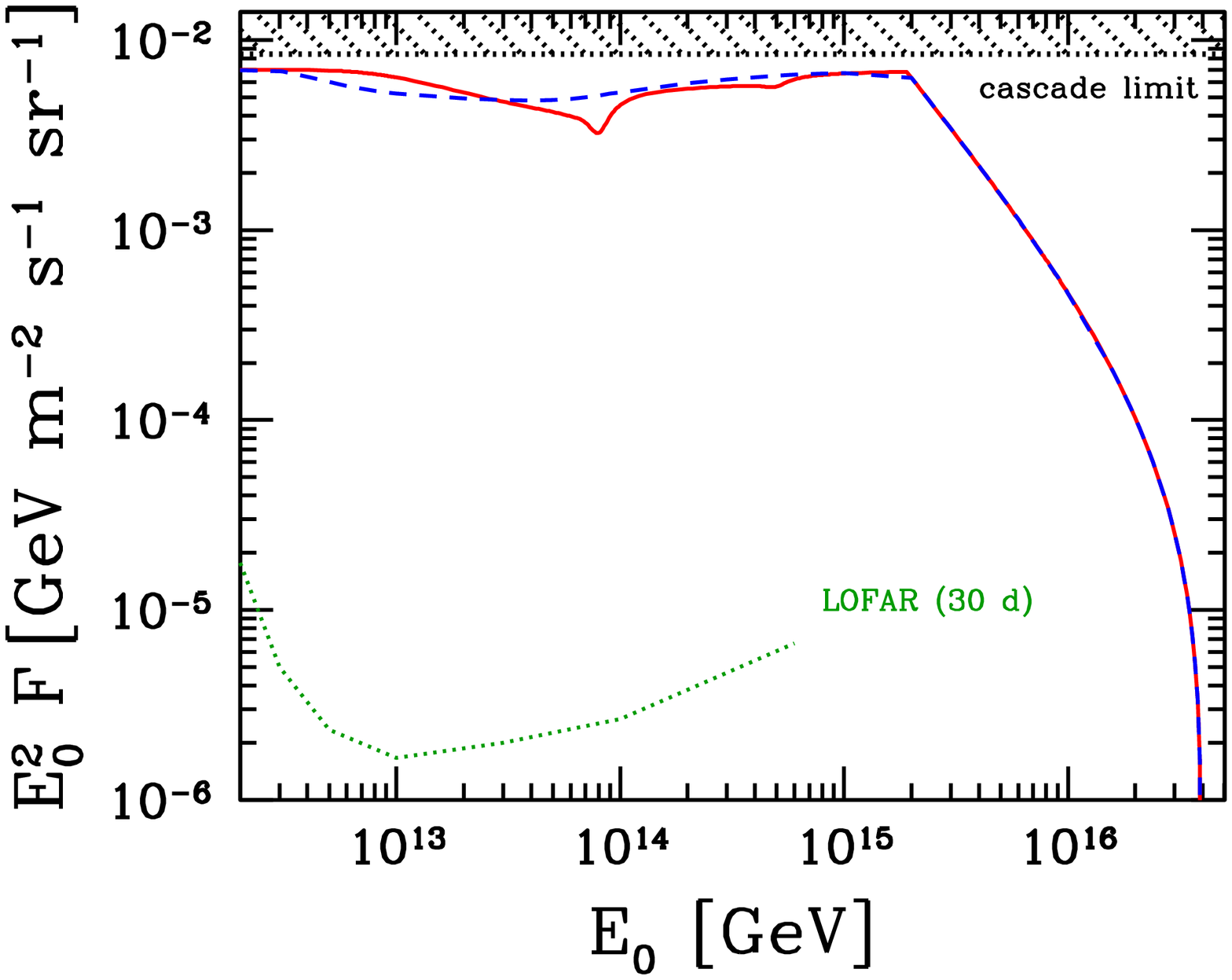}
\includegraphics*[bbllx=20pt,bblly=221pt,bburx=590pt,bbury=673pt,height=7cm,width=7.63cm]{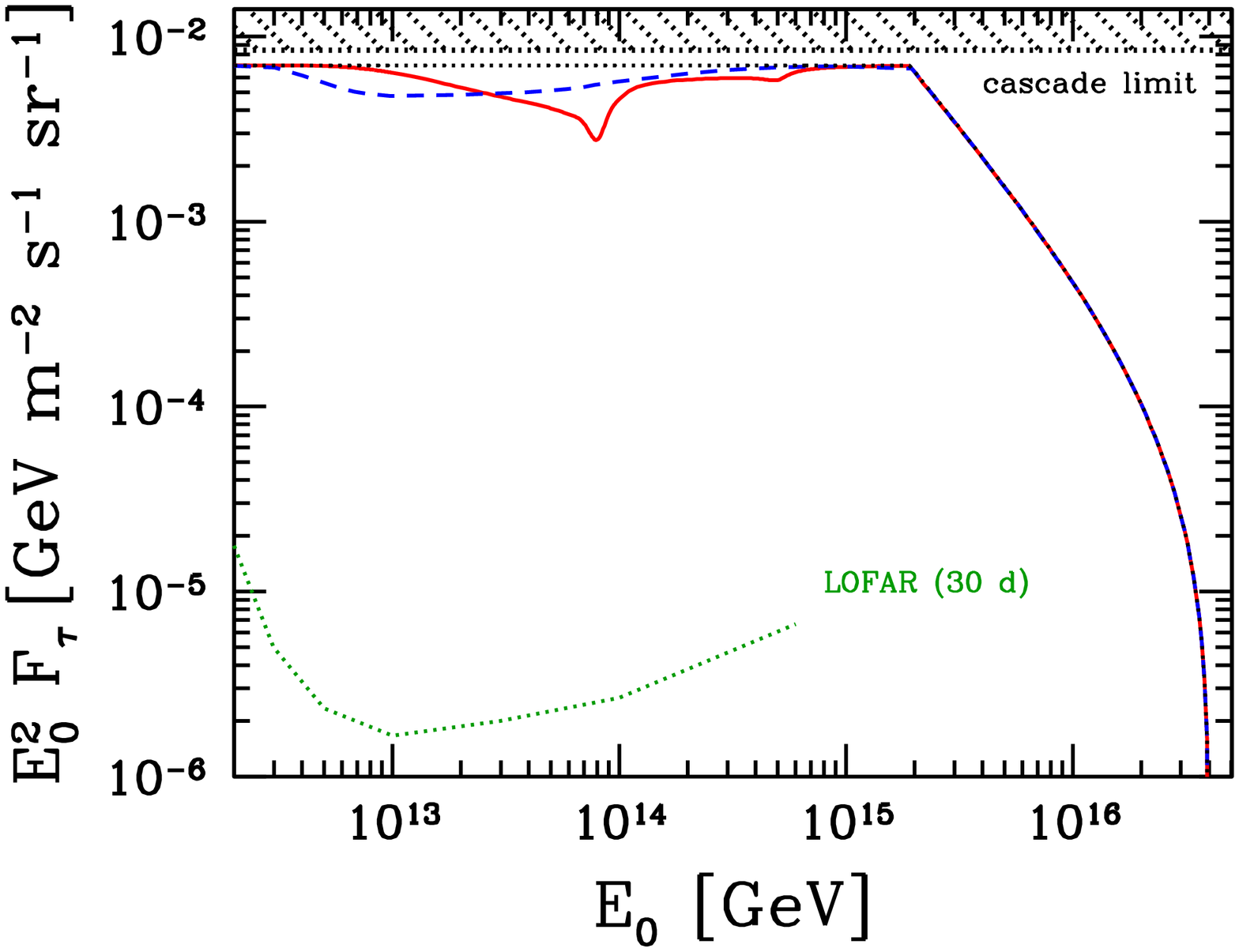}
\caption[]{Projected sensitivity of LOFAR~\cite{Scholten:2005pp} expressed in terms of the diffuse neutrino flux per flavor, corresponding to one event per energy decade and indicated duration, together with $E^2_0 F$ with $F=\sum F_{\nu_{\alpha}}+\sum F_{\bar{\nu}_{\alpha}}$ (left column) and $E^2_0 F_{\tau}$ with $F_{\tau}=F_{\nu_{\tau}}+F_{\bar{\nu}_{\tau}}$ (right column) for varying (solid lines) and constant (dashed lines) neutrino masses and for $z_{\max}=20$, assuming a normal neutrino mass hierarchy with $m_{\nu_{0,1}}=10^{-5}$ eV, $n=4$ and $\alpha=2$ as well as $E_{\max}=4\times 10^{16}$ GeV.} 
\label{fig6}
\end{center}
\end{figure}

For the purpose of illustrating our results, we consider Eq.~(\ref{eta}) and Eq.~(\ref{J}) for $n=4$ and $\alpha=2$ as often used in the literature to mimic astrophysical sources and take $E_{\max}=4\times 10^{16}$ GeV. As in the last sections, we assume a normal neutrino mass hierarchy according to Eq.~(\ref{m1}) -- Eq.~(\ref{m3}) and for the MaVaNs a neutrino mass variation according to Fig.~\ref{fig2}. We present our results in Fig.~\ref{fig6}, on the left hand side we plot the \UHEnu energy squared times the flavor summed flux $E^2_0 F$ with $F=\sum F_{\nu_{\alpha}}+\sum F_{\bar{\nu}_{\alpha}}$ according to Eq.~(\ref{FluxSum}) and on the right hand side $E^2_0 F_{\tau}$ with $F_{\tau}=F_{\nu_{\tau}}+F_{\bar{\nu}_{\tau}}$ as defined in Eq.~(\ref{Flux}). In Fig.~\ref{fig6} we plot our results both for varying (solid lines) and constant (dotted lines) neutrino masses for $z_{\max}=20$, together with the projected sensitivity of LOFAR~\cite{Scholten:2005pp} to be in operation by 2008 expressed in terms of diffuse fluxes per neutrino flavor, respectively. 

In our calculation we have assumed the \UHEnu flux to be close to the so-called cascade limit~\cite{B,Mannheim:1998wp}. It applies to sources where the neutrinos emerge from pion decays or even from electroweak jets~\cite{Berezinsky:2002hq} and are thus accompanied by photons and electrons which escape the source. Consequently, the measurements of diffuse gamma-ray fluxes, which are of the same origin but have cascaded down in energy during the propagation through the universe, restrict the neutrino flux to lie below the cascade limit. Apparently, the predicted sensitivity of LOFAR~\cite{Scholten:2005pp}, corresponding to one event per neutrino flavor per energy decade, lies below the cascade limit by several orders of magnitude. Accordingly, at best 3500 neutrinos (plus anti-neutrinos) in the energy interval $10^{12}-10^{13}$ GeV can be expected to be detected by the radio telescope.

\begin{figure}
\begin{center}
\includegraphics*[bbllx=20pt,bblly=101pt,bburx=590pt,bbury=658pt,height=6.8cm,width=7.7cm]{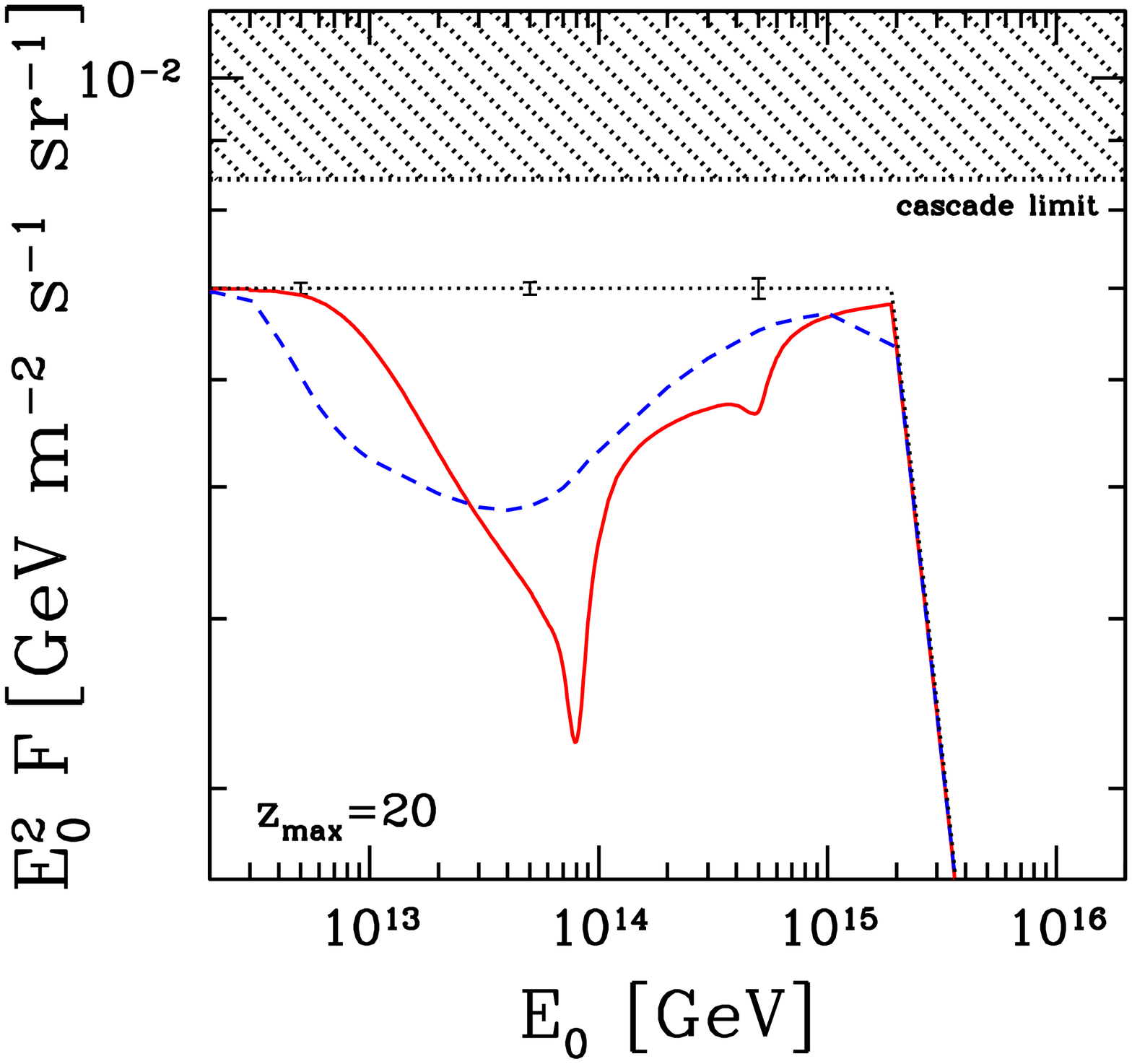}
\includegraphics*[bbllx=20pt,bblly=101pt,bburx=590pt,bbury=658pt,height=6.8cm,width=7.7cm]{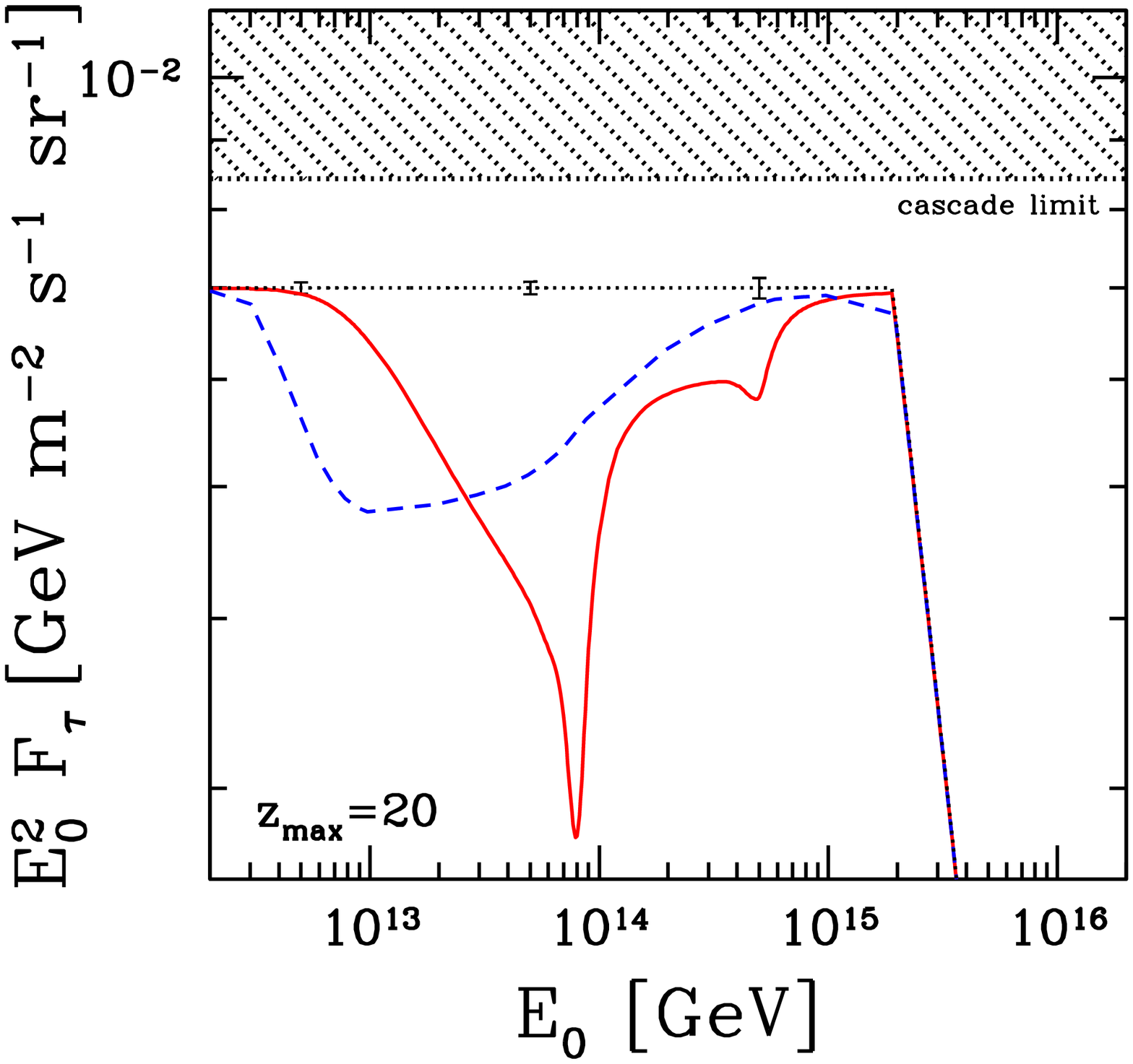}
\includegraphics*[bbllx=20pt,bblly=101pt,bburx=590pt,bbury=658pt,height=6.9cm,width=7.7cm]{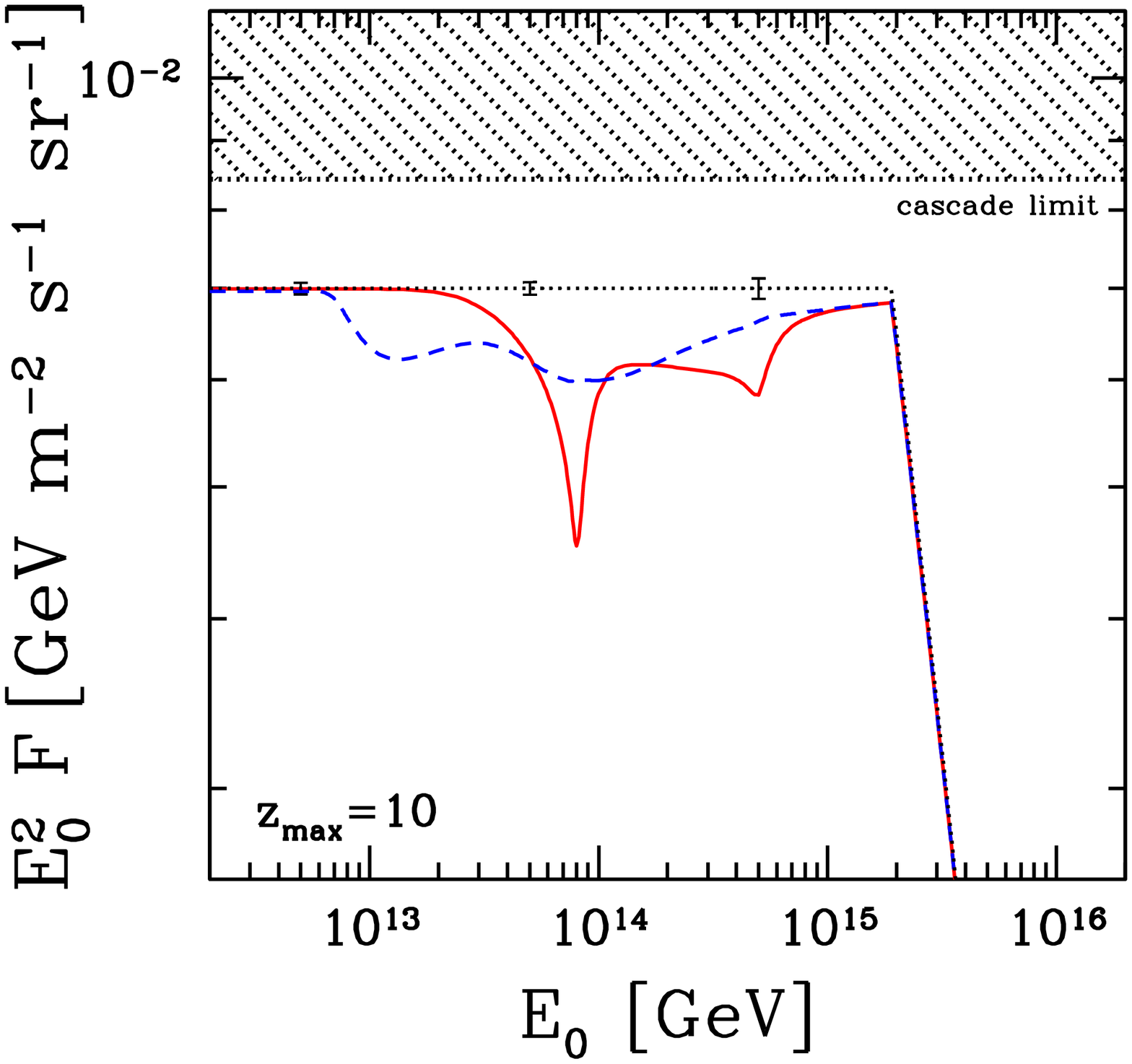}
\includegraphics*[bbllx=20pt,bblly=101pt,bburx=590pt,bbury=658pt,height=6.9cm,width=7.7cm]{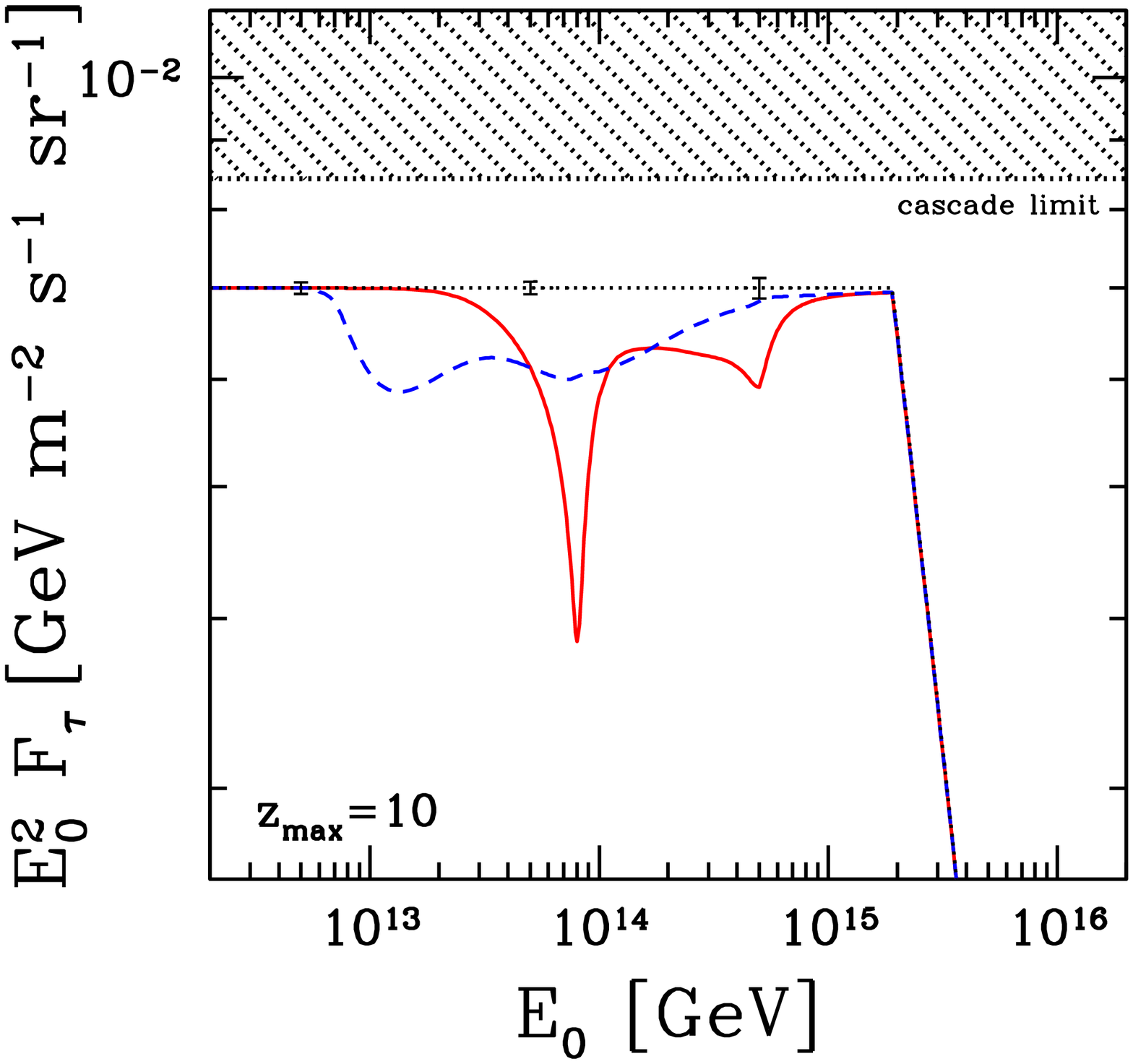}
\includegraphics*[bbllx=20pt,bblly=101pt,bburx=590pt,bbury=658pt,height=6.9cm,width=7.7cm]{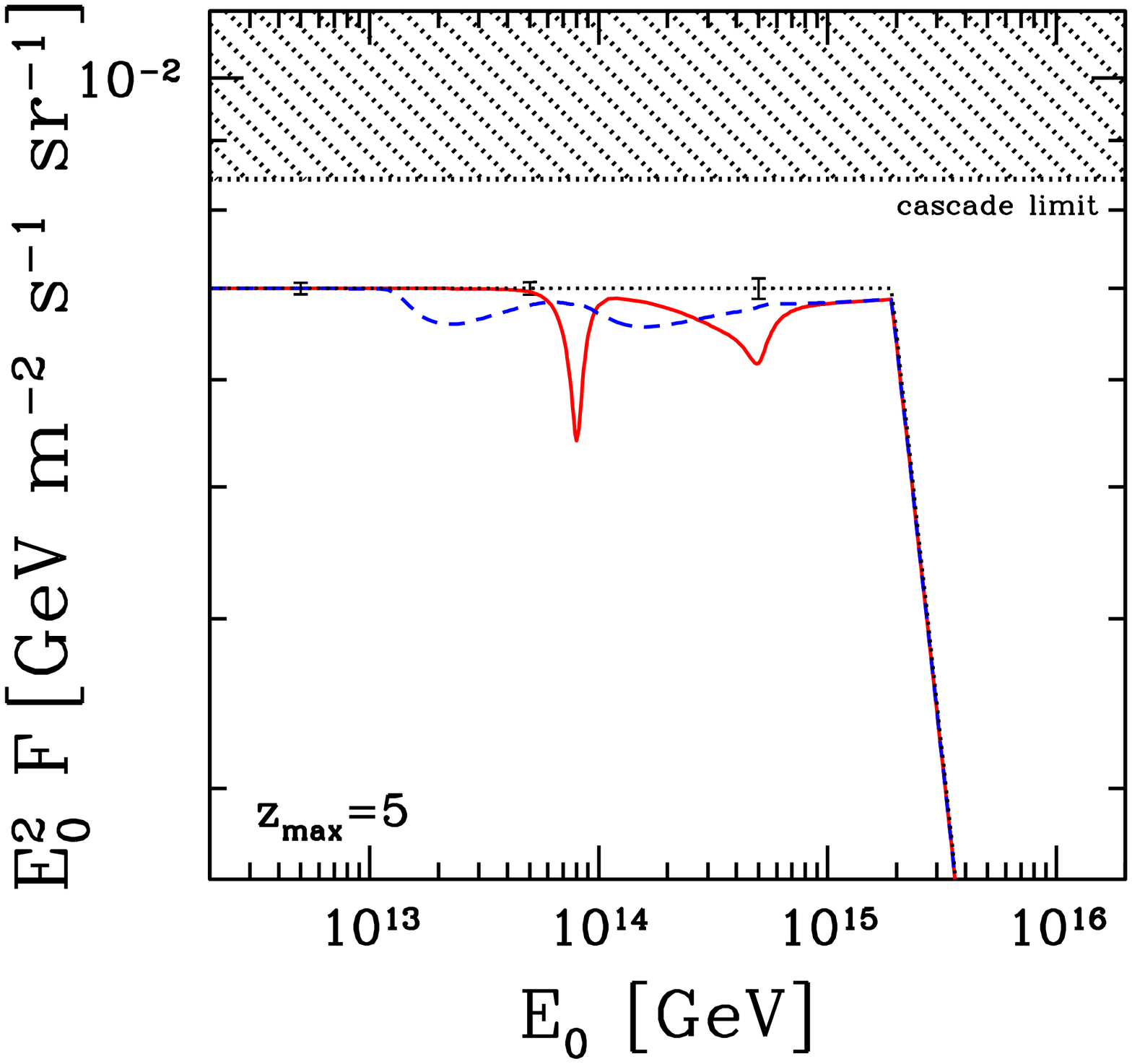}
\includegraphics*[bbllx=20pt,bblly=101pt,bburx=590pt,bbury=658pt,height=6.9cm,width=7.7cm]{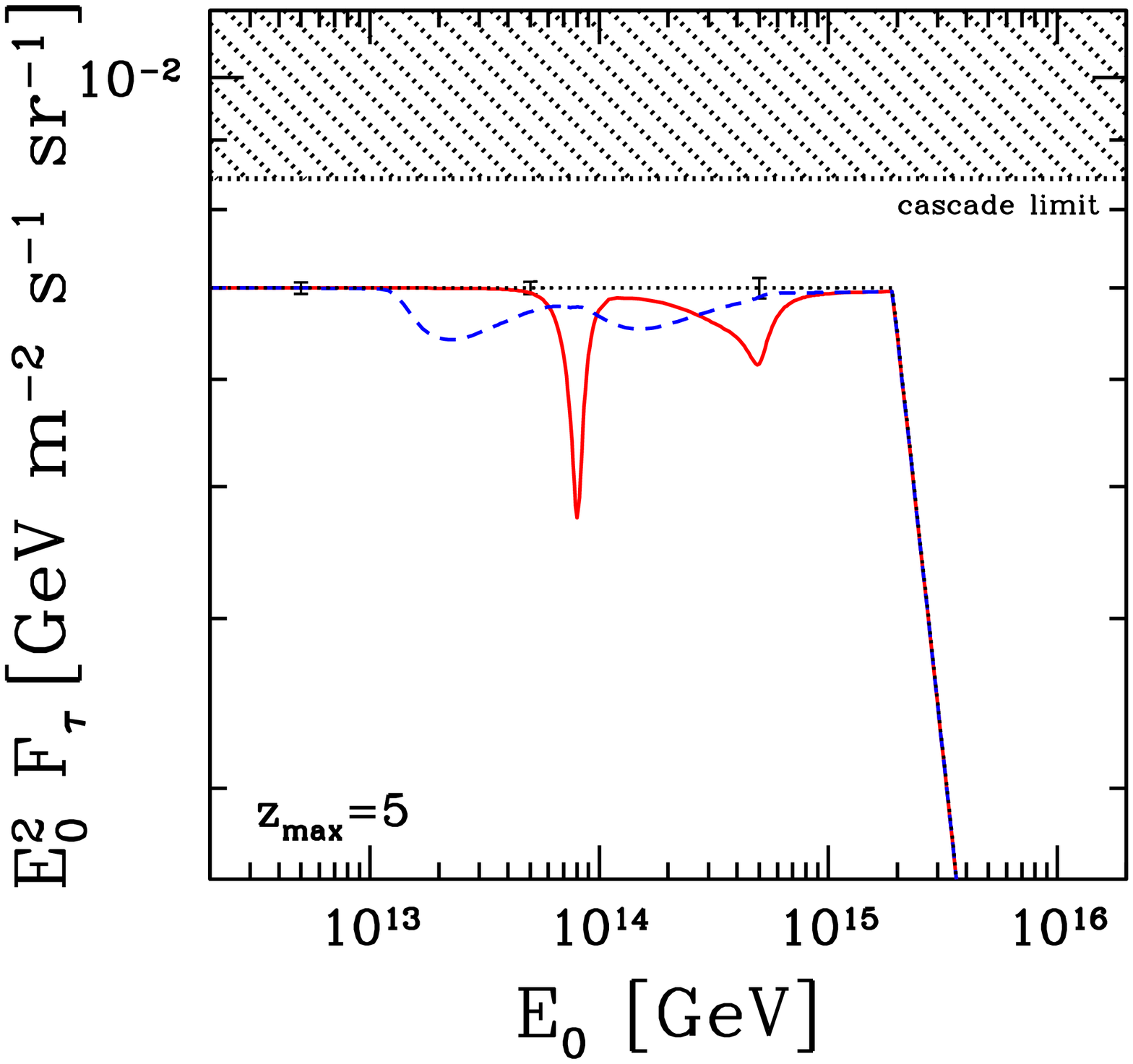}
\caption[]{The energy squared times the flavor summed neutrino flux $E^2_0 F$ with $F=\sum F_{\nu_{\alpha}}+\sum F_{\bar{\nu}_{\alpha}}$ (left column) and $E^2_0 F_{{\tau}}$ with $F_{\tau}=F_{\nu_{\tau}}+F_{\bar{\nu}_{\tau}}$ (right column) for varying (solid lines) and constant (dotted lines) neutrino masses for $z_{\max}=20$, $z_{\max}=10$ and $z_{\max}=5$ from top to bottom, respectively. All curves assume a normal neutrino mass hierarchy with $m_{\nu_{0,1}}=10^{-5}$ eV, $n=4$ and $\alpha=2$ as well as $E_{\max}=4\times 10^{16}$ GeV.} 
\label{fig7}
\end{center}
\end{figure}

The large event numbers $N$ result in tiny error bars ($\sigma=\sqrt{3N}/3N$). In a blow-up of the absorption features in Fig.~\ref{fig7}, we have adjusted them to the curves with no absorption for emission redshifts $z_{\max}=20$, $z_{\max}=10$ and $z_{\max}=5$ from top to bottom. Thereby, we have assumed a conservative and therefore rather poor energy resolution corresponding to one energy bin per energy decade, whereas at best LOFAR is predicted to achieve an energy resolution of $\Delta E/E\sim 30\%$~\cite{Private}. The latter would correspond to $\sim 4$ energy bins per energy decade. Apparently, the dips become considerably deeper with increasing $z_{\max}$. Despite the underlying low neutrino mass scale, both for varying and constant neutrino masses, LOFAR can be expected to produce significant evidence for absorption dips in the \UHEnu spectra for emission redshifts $z_{\max}=20$ and $z_{\max}=10$ -- even for a bad energy resolution. In the case of $z_{\max}=5$, in the interval $10^{14}-10^{15}$ GeV, the considerably higher dip depth for MaVaNs leads to a more than $5\sigma$ deviation from the curve with no absorption whereas for constant mass neutrinos the departure is not significant (both for the flavor summed flux $F$ and for $F_{\tau}$). Even if the underlying \UHEnu fluxes are much lower, at least for \UHEnu sources at $z_{\max}=20$, a detection of absorption features produced by varying, light neutrino masses could well be feasible. Correspondingly, if such \UHEnu fluxes of astrophysical origin exist, the most direct detection of the \CnuB so far seems to be in reach within the next decade. From the experimental point of view, the prospects are even better for scenarios with time varying neutrino masses, which in general produce deeper absorption dips in the regime of astrophysical emission redshifts. 

Let us now turn in more detail to the prospects of probing scenarios of Neutrino Dark Energy by identifying the characteristic absorption signatures of a possible neutrino mass evolution. Since the \UHEnu fluxes are governed by the respective survival probabilities discussed in the last section, the characteristic differences in the absorption features for varying instead of constant neutrino masses are maintained. Namely, for MaVaNs one observes a clear shift of the dips to higher energies as well as considerably deeper absorption minima with respect to constant mass neutrinos. Accordingly, given a decent energy resolution of $\Delta E/E\sim 30\%$~\cite{Private} for LOFAR, relic neutrino absorption spectroscopy could serve as a test for the nature of neutrino masses and therefore for Neutrino Dark Energy. However, certainly, the feasibility strongly depends on the energy resolution achieved by the \UHEnu observatory. 

\subsection{Topological defect neutrino sources}                                                                           %

In the following we will discuss neutrino fluxes expected to result from exotic top-down \UHEnu sources like topological defects. As already mentioned, \UHEnu's might be produced among other Standard Model particles in the decays of super-heavy $X$ quanta which constitute the topological defects. Accordingly, the corresponding \UHEnu injection spectra $J$ in Eq.~(\ref{LumFac}) are fragmentation functions which can reliably be predicted by the help of Monte Carlo generators~\cite{Birkel:1998nx,Berezinsky:2000up} or via the Dokshitzer-Gribov-Lipatov-Altarelli-Parisi (DGLAP) evolution~\cite{Fodor:2000za,Sarkar:2001se,Barbot:2002gt} from experimentally determined initial distributions at a scale $M_Z$ to the ones at $m_X$. The corresponding injection rate (the activity $\eta$ in Eq.~(\ref{LumFac})), which in particular determines the overall normalization of the neutrino flux, in terms of cosmic time $t$ is given by,
\be
\label{injectionrate}
\frac{\partial n_X}{\partial t}= \frac{Q_0}{m_X} \left(\frac{t}{t_0}\right)^{-4+p},
\ee
where $Q_0$ is the energy emitted per unit volume per unit time at present and $p$ is a dimensionless constant. Both $Q_0$ and $p$ depend on the specific topological defect scenario~\cite{Bhattacharjee:1991zm}. In our analysis, we calculate the absorption features in the \UHEnu spectra for topological defect sources by the help of fragmentation functions as well as by Eq.~(\ref{injectionrate}) according to Eq.~(\ref{LumFac}).\footnote{In the literature on absorption dips so far the injection spectra of top-down sources had been approximated by Eq.~(\ref{eta}) and Eq.~(\ref{J}). Cf. e.g. Refs.~\cite{Eberle:2004ua,D'Olivo:2005uh} for the appropriate values for $n$ and $\alpha$.} 

\begin{figure}
\begin{center}
\includegraphics*[bbllx=20pt,bblly=191pt,bburx=570pt,bbury=673pt,height=7.0cm,width=7.7cm]{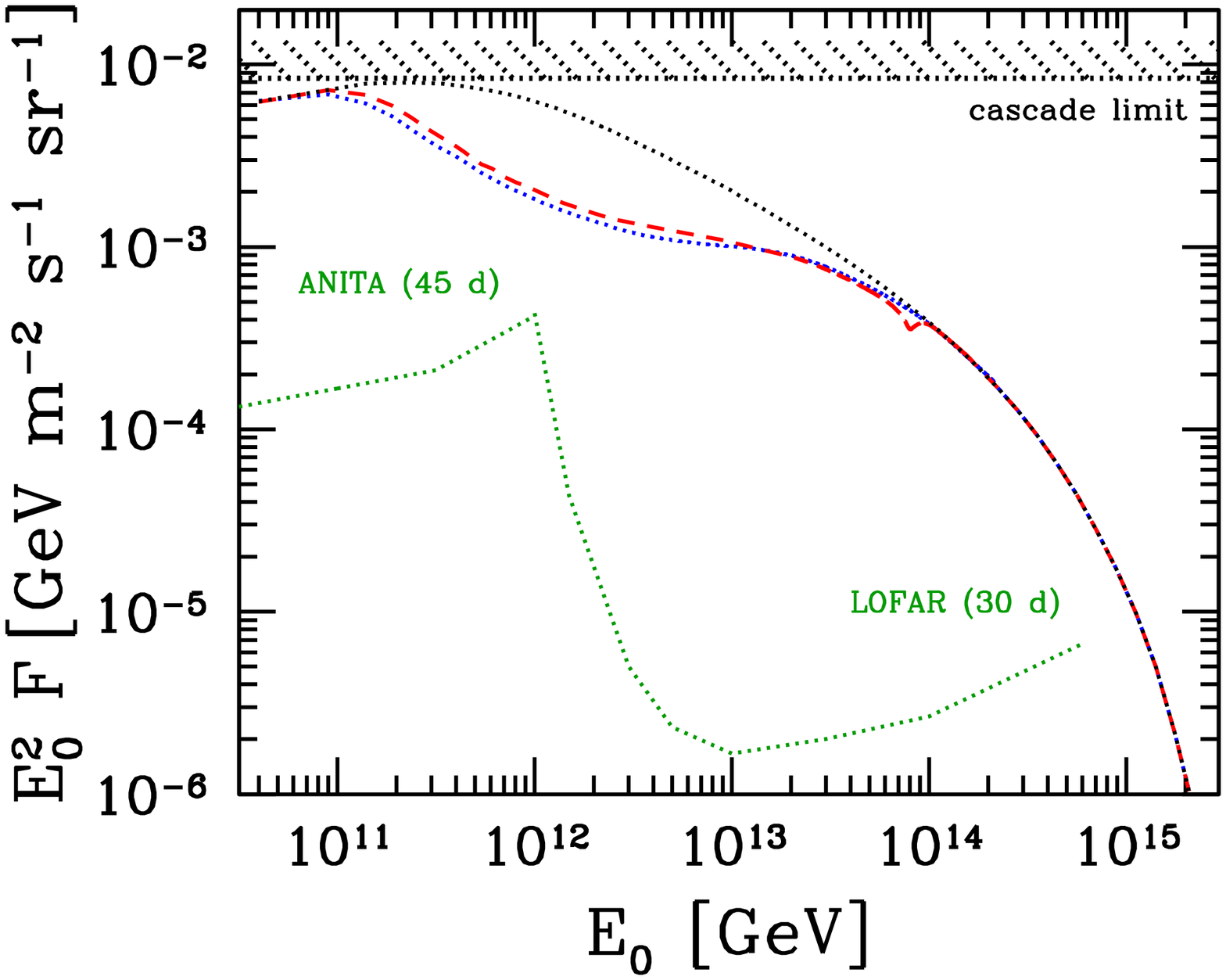}
\includegraphics*[bbllx=20pt,bblly=191pt,bburx=570pt,bbury=673pt,height=7.0cm,width=7.7cm]{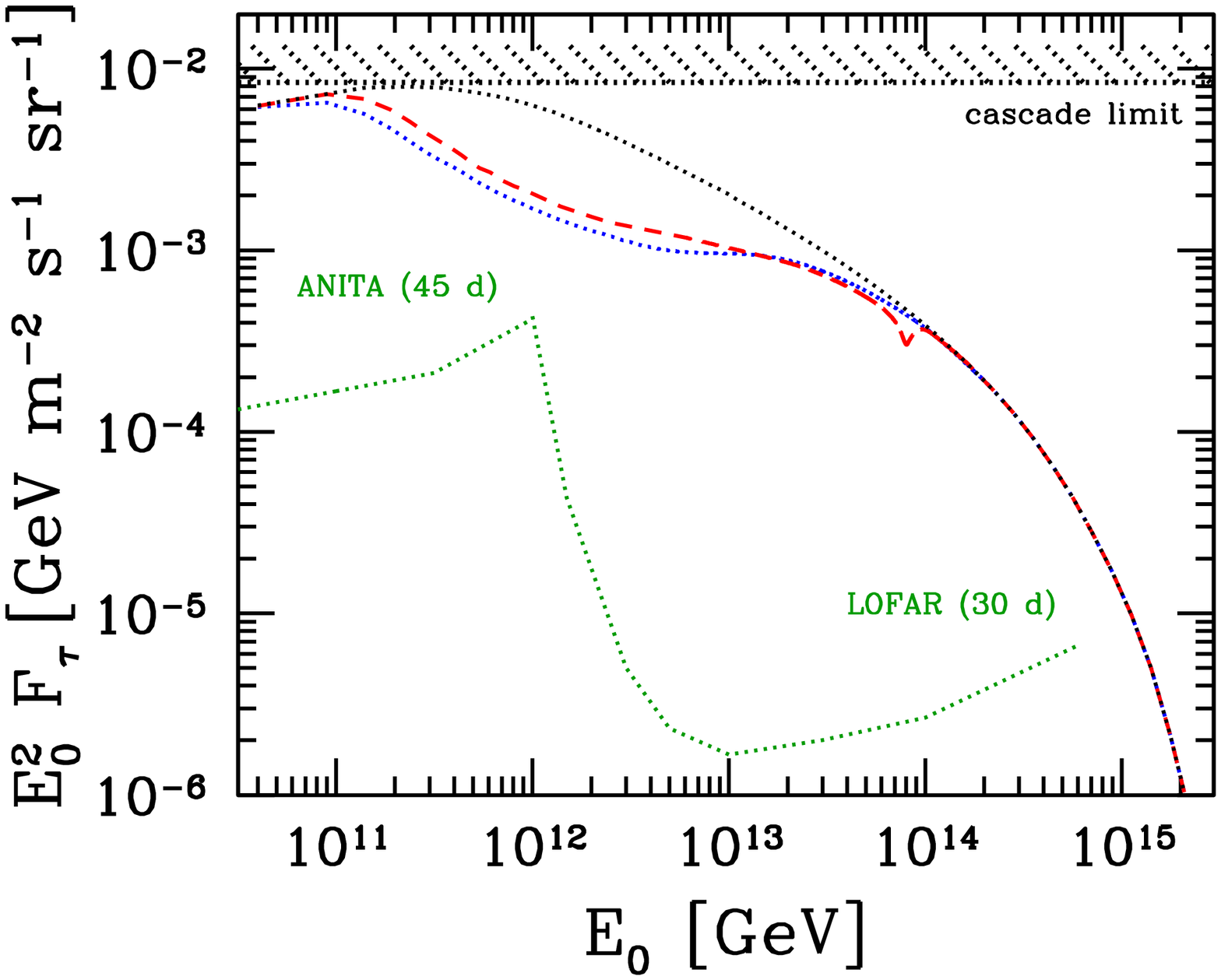}
\includegraphics*[bbllx=20pt,bblly=221pt,bburx=570pt,bbury=673pt,height=6.6cm,width=7.7cm]{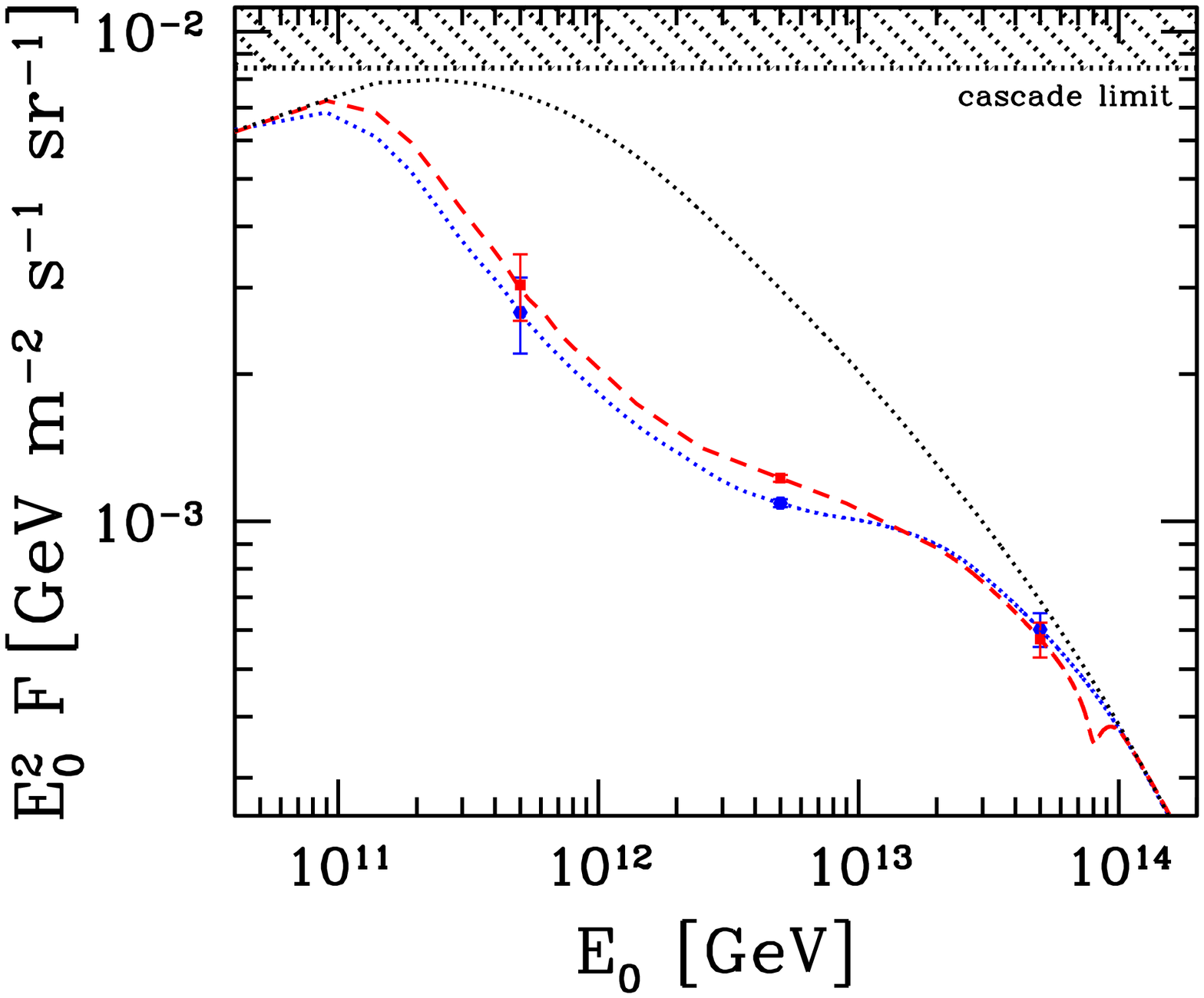}
\includegraphics*[bbllx=20pt,bblly=221pt,bburx=570pt,bbury=673pt,height=6.6cm,width=7.7cm]{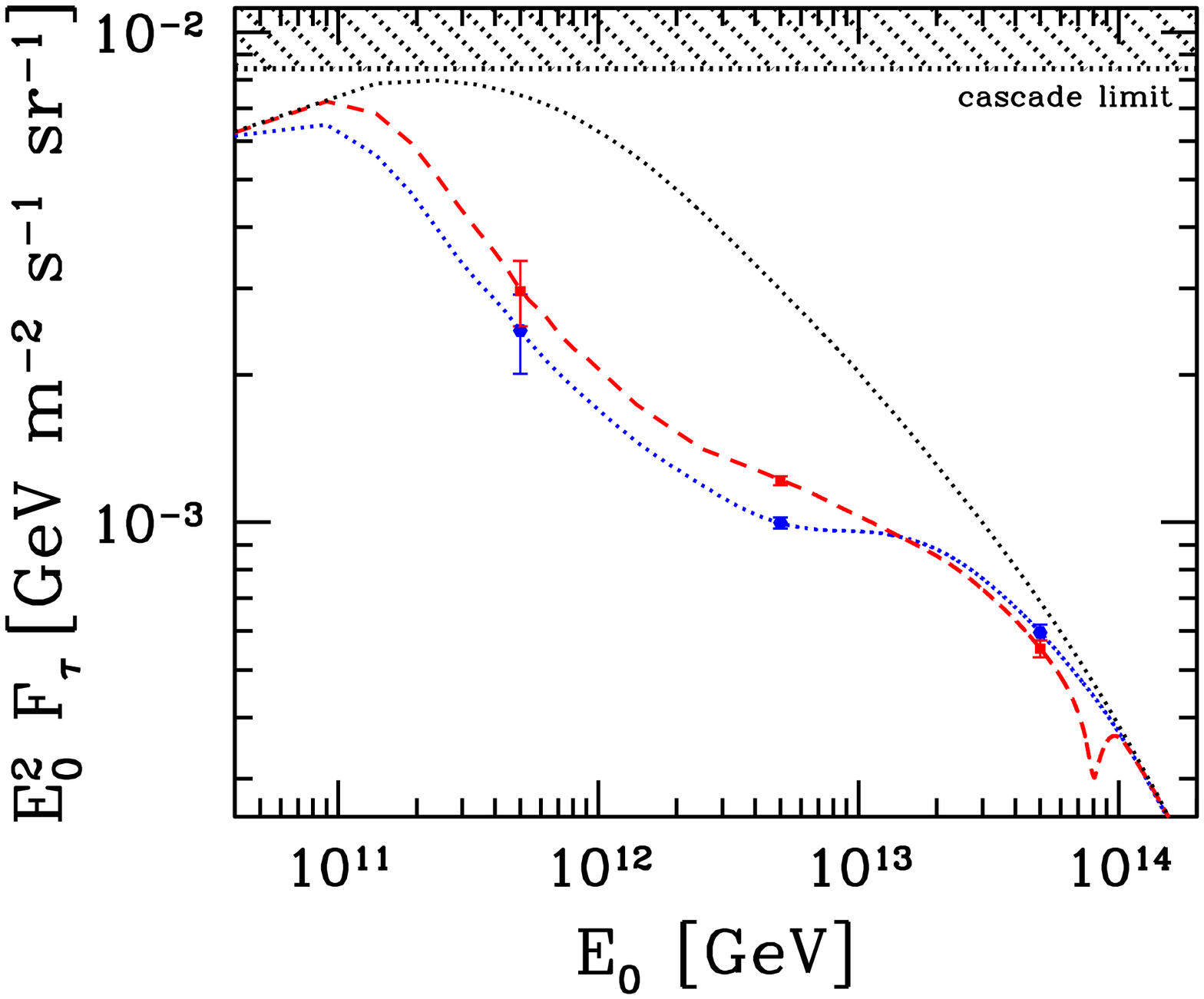}
\caption[]{The energy squared times the neutrino flux $E^2_0 F$ with $F=\sum F_{\nu_{\alpha}}+\sum F_{\bar{\nu}_{\alpha}}$ (left column) and $E^2_0 F_{\tau}$ with $F_{\tau}=F_{\nu_{\tau}}+F_{\bar{\nu}_{\tau}}$ (right column) for varying (dashed lines) and constant (dotted lines) neutrino masses expected from the decomposition of a superconducting string with $p=0$ and $m_X=10^{16}$ GeV, in the first panel together with the projected sensitivities for ANITA~\cite{Barwick:2005hn} and LOFAR\cite{Scholten:2005pp}, which correspond to one event per flavor, energy decade and indicated duration, respectively.}
\label{fig8}
\end{center}
\end{figure}

In the following, we will illustrate the prospects for performing relic neutrino absorption spectroscopy for top-down sources considering as example superconducting strings whose evolution is characterized by $p=0$ in Eq.~(\ref{injectionrate}). Note that in the energy region of the absorption dips the Standard Model and SUSY fragmentation functions (which we have taken from~\cite{Sarkar:2001se}) have practically the same shape. Accordingly, all our results on relic neutrino absorption are independent of the supersymmetrization of the MaVaN scenario (cf. Sec.~\ref{sec:MaVaNs}). 

In Fig.~\ref{fig8} we plot the expected absorption features for varying (dashed lines) and constant (dotted lines) neutrino masses as defined in Eq.~(\ref{m1})-Eq.~(\ref{m3}), where for the MaVaNs we assume a mass variation according to Fig.\ref{fig2}. Again, we present our results for the energy squared times the flavor summed flux $E^2_0 F$ according to Eq.~(\ref{FluxSum}) in the left column of Fig.~\ref{fig8} and $E^2_0 F_{\tau}$ as defined in Eq.~(\ref{Flux}) in the right column of Fig.~\ref{fig8}. As in the case of astrophysical sources, in the first panel of each column we have plotted the projected sensitivities of ANITA and LOFAR as well as the predicted \UHEnu flux for $m_X=10^{16}$ GeV with and without absorption (where the latter by design scratches the cascade limit). In the second panel of Fig.~\ref{fig8}, we again show a blow-up of the absorption features. Apparently, with respect to the astrophysical sources (cf. Fig.~\ref{fig7}), the dips are broader by almost an order of magnitude in energy. This is due to the fact that the constituents of topological defects have started to decay and therefore to release neutrinos at $z\gg 1$. As a further consequence, the dips for top-down sources are much deeper than for bottom-up sources, since the survival probability of a neutrino traveling to us is much lower (as discussed in the last section and as illustrated by the comparison of the flavor summed survival probability for $z_s=50$ and $z_s=20$ in Fig.~\ref{fig5}). Consequently, both of these features facilitate a detection of the absorption dips in the \UHEnu spectra. Accordingly, top-down sources with the same underlying fluxes as astrophysical sources are even better suited to provide evidence for the existence of the \CnuB. By this means valuable information on the topological defect scenario, on cosmological parameters as well as the neutrino mass scale could be gained both for varying and constant neutrino masses. Furthermore, for the first time in cosmic particle physics, the GUT energy scale ${\cal{O}}(m_X)\sim 10^{16}$ GeV could be probed.  

In the second panel of Fig.~\ref{fig8}, we have also included the expected error bars, again assuming one energy bin per energy decade, both for the MaVaN and constant mass neutrino absorption lines. Clearly, the discrepancy between the two curves is larger for $E^2_0 F_{\tau}$ than for $E^2_0 F$, whereby in contrast to \UHEnu's of astrophysical origin, constant neutrino masses produce somewhat deeper dips than time dependent masses. These features can be understood by realizing that $\nu_{\tau}$ is dominantly composed of the heaviest mass eigenstate $m_{\nu_3}$, whereas $E^2_0 F$ by definition gets equal contributions from all mass eigenstates. Furthermore, for constant mass neutrinos, $m_{\nu_3}$ is the only mass eigenstate for which the ratio $m_{\nu_3}/T_{\nu}(z)\sim (1+z)\gg 1$ up to $z \sim 1000$. In other words, it produces much deeper absorption dips than the lighter mass eigenstates (even when integrating back to $z\gg 1$) and their characteristic shape is not dominated by the temperature effects (cf. the discussion in the last section). In contrast, for MaVaNs, the ratio $\mnui(z)/T_{\nu}(z)$ with $i=1,2,3$ for all mass eigenstates drops much faster with increasing $z$ and takes values $\mnui(z)/T_{\nu}(z)\ll 1$ for $z\gg 1$. As a result, the low energy end of the dip (which corresponds to higher annihilation redshifts $z$) has the same shape both for all MaVaN mass eigenstates and the lighter two constant ones and is totally determined by the thermal background effects. 

In summary, promisingly, a resolution of absorption features for either mass behavior seems to be possible both for ANITA and for LOFAR. Yet, a differentiation of the MaVaN and constant mass neutrino absorption features seems only feasible, if tau flavor tagging and a good energy resolution are achieved.

\section{Summary and Conclusions \label{sec:Conclusions}} %

In light of the number of extremely high-energy neutrino (\UHEnu) observatories in operation and under construction with a combined sensitivity ranging up to $10^{17}$ GeV, the prospects for establishing the existence of \UHEnu fluxes appear to be very promising. As a next step, the exciting possibility opens up to trace the annihilation of \UHEnu's and relic anti-neutrinos (and vice versa) into $Z$ bosons by localizing absorption dips in the \UHEnu spectra at energies set by the neutrino masses. On the one hand, their detection could furnish the most direct evidence for the \CnuB so far and thereby confirm standard cosmology back to the time of light neutrino decoupling. On the other hand, the shape of the absorption lines could reveal a variation of neutrino masses with time and thus verify the interpretation of the \CnuB as source of Neutrino Dark Energy.

We therefore considered a viable Mass Varying Neutrino (MaVaN) model with the following features entering our analysis on relic neutrino absorption. By the requirement that the lightest neutrino still has to be moderately relativistic today the neutrino mass scale is set to be low, which leads to very conservative predictions. Furthermore, the evolving neutrino masses $\mnui(z)$, which we determined numerically as functions of redshift assuming $m_{\nu_{0,1}}=10^{-5}$ eV, turned out to be well approximated by simple power laws $(1+z)^{-1}$ and $(1+z)^{-1/2}$ in the low and in the high redshift regime, respectively. Accordingly, as a generically important feature, they are decreasing functions of redshift as in all standard MaVaN scenarios. 

In order to provide all technical tools to interpret \UHEnu absorption dips for a given injection spectrum and to extract valuable information on neutrino physics, cosmology and possibly physics beyond the Standard Model, we proceeded in the following way. We considered in parallel the neutrino masses to be functions of cosmic time as well as to be constants. In our analysis we took into account the full thermal background effects which result from the relic neutrino motion according to their phase space distribution. In order to compare our results to the literature, we included in our discussion common approximations~\cite{Roulet:1992pz,Eberle:2004ua,Barenboim:2004di} which neglect part or all of the dependence of the damping on the relic neutrino momenta.  

On the level of the survival probabilities which govern the \UHEnu fluxes, we found the following results: For low emission redshifts ($z\sim {\cal{O}}(5)$), the absorption dips produced by the varying neutrino masses $\mnui(z)$ for $i=2,3$ exhibit narrow absorption minima, which do not suffer a distortion to lower energies as the corresponding dips of constant mass neutrinos. As a consequence, for MaVaNs, the absorption dips of the flavor components $\nu_{\mu}$ and $\nu_{\tau}$ (which are mostly composed of the heavier two mass eigenstates) are clearly deeper and shifted to higher energies by almost an order of magnitude with respect to the corresponding constant mass minima. For an increased emission redshift $z\gg 5$, these features become somewhat less pronounced but essentially prevail. A better understanding of the characteristic signatures caused by the mass evolution was obtained by switching off the superposing thermal wash-out caused by the relic neutrino motion. After neglecting the relic neutrino momenta for this purpose, we found that the crucial deviations result from the dependence of the corresponding resonance energies on the neutrino masses $E^{\res}_{i}\sim 1/ \mnui$ for $i=1,2,3$. In the case of MaVaNs, the mass variation $\mnui(z)$ induces a dependence on the annihilation redshift $z$ according to $E^{\res}_{i}(z)\sim E^{\res}_{0,i}(1+z)$ for all neutrino species $i=1,2,3$ in the low redshift regime. Accordingly, this $z$ dependence of the resonance energies compensates for the energy loss of the \UHEnu due to cosmic redshift proportional to $(1+z)^{-1}$, resulting in narrow absorption spikes at constant energies $E^{\res}_{i}(z)/(1+z)=E^{\res}_{0,i}$ (like one would expect for constant neutrino masses in a non-expanding universe). In contrast, for constant neutrino masses $\mnui=m_{\nu_{0,i}}$ the absorption dips are broadened, since the redshifted resonance energies to be measured on earth are given by $E^{\res}_{i}/(1+z)=E^{\res}_{0,i}/(1+z)$, for $z$ taking values between $0$ and the \UHEnu emission redshift. Since in the standard MaVaN scenario the neutrino masses are decreasing functions of redshift, they generically reduce the effect of cosmic redshift on the \UHEnu survival probabilities. As a result, they always produce deeper absorption minima, which, in addition, are shifted to higher energies in comparison to the dips caused by constant neutrino masses.

In order to illustrate the discovery potential for absorption dips in the \UHEnu spectra to be observed at earth and to estimate the prospects of testing scenarios of Neutrino Dark Energy, we considered plausible \UHEnu fluxes originating from astrophysical acceleration sites or from topological defect sources. We presented our results both for the energy squared times the flavor summed \UHEnu flux $E^2_0 F$  with $F=\sum F_{\nu_{\alpha}}+\sum F_{\bar{\nu}_{\alpha}}$ and for $E^2_0 F_{\tau}$ with $F_{\tau}=F_{\nu_{\tau}}+F_{\bar{\nu}_{\tau}}$, where the latter can at best be identified by LOFAR~\cite{Private}. Despite the adopted low neutrino mass scale, we found both for varying and constant neutrino masses that for topological defect and for astrophysical \UHEnu sources at $z_{\max}>5$, LOFAR and ANITA promise a statistically significant evidence for absorption dips (even if the underlying fluxes are well below the cascade limit). Accordingly, the most direct detection of the \CnuB so far seems to be in reach within the next decade. 

Furthermore, the flux dips of varying and constant mass \UHEnu's expected from astrophysical sources retain the characteristic differences induced by the survival probabilities. Besides being clearly shifted to higher energies, the MaVaN dips are deeper and therefore even facilitate a resolution of absorption features in the \UHEnu spectra in comparison to constant mass neutrinos both in the case of $E^2_0 F$ and of $E^2_0 F_{\tau}$. As a main result of our analysis, these deviations of the MaVaN and constant mass absorption curves for astrophysical sources turned out to be statistically significant, yet a decent energy resolution seems necessary for their detection. Given an energy resolution of $\Delta E/E\sim 30\%$ as at best achievable for LOFAR~\cite{Private}, relic neutrino absorption spectroscopy could reveal a variation of neutrino masses and therefore possibly the nature of Dark Energy. 

As concerns topological defect sources, the absorption lines in the \UHEnu fluxes for time dependent and constant neutrino masses altogether are more similar in shape, however, somewhat deeper for constant neutrino masses. Furthermore, they extend to much lower energies than for astrophysical \UHEnu sources and their minima are considerably deeper. All of these features are a result of the much higher annihilation redshifts $z_s\gg 1$ possible for \UHEnu's originating from the decomposition of topological defects in comparison to \UHEnu's from astrophysical acceleration sites. At high redshifts, the \UHEnu's are absorbed by a hotter bath of relic neutrinos. Consequently, in the energy region spanned by the absorption dips where $\mnui/T_{\nu} \ll 1$, thermal background effects wash out any features produced by the neutrino mass or its possible variation. Since the MaVaN masses are decreasing functions of redshift, they reach this limit for much smaller redshifts than the corresponding constant masses. Only the mass of the heaviest constant mass eigenstate is sufficiently large, $m_{\nu_3}/T_{\nu}(z)\gg 1$, in the relevant energy region, leading to a deeper absorption curve than the one produced by all of the other MaVaN and constant mass eigenstates. Since ${\nu_{\tau}}$ is mostly composed of the heaviest mass eigenstate, $F_{\tau}$ exhibits deeper constant mass dips than $F$. Accordingly, for $F_{\tau}$ the signatures of varying neutrino masses can more easily be distinguished from those of constant masses than in the case of $F$. However, in order to reveal a neutrino mass variation, it seems necessary both to identify the tau neutrino flavor and to have a good energy resolution.

Recently, the authors of Ref.~\cite{Takahashi:2006jt} claimed that certain constraints on the fundamental scalar acceleron potential $V_0(\A)$ (cf. Sec.~\ref{sec:MaVaNs}) can also lead to MaVaN models stable against the growth of inhomogeneities~\cite{Afshordi:2005ym} even in the highly non-relativistic regime. Accordingly, one could construct other MaVaN models than the one under consideration in this paper, whose viability would not rely on a low neutrino mass scale (cf. Sec.~\ref{sec:MaVaNs}). As we pointed out, the characteristic absorption signatures of any standard MaVaN scenario (cf. Sec.~\ref{sec:SurvivalProb} and Sec.~\ref{sec:AbsorptionDips}) are essentially generic apart from details. Yet, a higher neutrino mass scale would even increase the overall dip depth in comparison to our rather conservative predictions and also reduce the importance of the thermal background effects on the absorption features. Accordingly, we would expect the features produced by a possible mass evolution not to be washed out by the temperature effects for a wider energy range of the dips. Thus the deviations with respect to the corresponding constant mass curves would even be more prominent.   

\section*{\bf Acknowledgments} %
We thank Yvonne Wong for fruitful discussions and important information, and Lily Schrempp thanks her for technical advice. Information about LOFAR from Jose Bacelar is also kindly acknowledged. Furthermore, we thank Markus Ahlers, Joerg Jaeckel and Joern Kersten for helpful conversations, and Lily Schrempp thanks Markus Ahlers for technical advice.   

\section*{References}
\frenchspacing
\bibliography{JCAP}
\addcontentsline{toc}{section}{Bibliographie}

\bibliographystyle{utcaps}

\end{document}